\documentclass[useAMS,usenatbib]{mn2e}

\input{epsf}

\usepackage{graphicx}	
\usepackage{deluxetable}
\usepackage{amsmath}

\newcommand{\CIVw}{C\,{\sc iv}~$\lambda$1549}

\newcommand{\OIIIw}{[O$\,\textsc{iii}]$~$\lambda$5007}

\newcommand{\CIV}{C\,{\sc iv}}

\newcommand{\Hb}{{H}\,$\beta$}
\newcommand{\FeII}{Fe\,\textsc{ii}}

\newcommand{\OIII}{[O\,{\sc iii}]}
\newcommand{\OIV}{O\,{\sc iv}]}

\newcommand{\SiIV}{Si\,{\sc iv}}

\newcommand{\siivoiv}{Si\,{\sc iv}+O\,{\sc iv}]}
\newcommand{\siivoivw}{Si\,{\sc iv}+O\,{\sc iv}]~$\lambda$1400}

\newcommand{\fwhmhb}{$\rm{FWHM}_{\rm{H}\beta}$}
\newcommand{\fwhmhbpred}{$\rm{FWHM}_{\rm{H}\beta,\rm{predicted}}$}

\newcommand{\fwhmciv}{FWHM$_{\rm{C\,{\sc IV}}}$}

\newcommand{\siglciv}{$\sigma_{l,\rm{C {\sc IV}}}$}

\def\lsim{\lower0.3em\hbox{$\,\buildrel <\over\sim\,$}}
\def\gsim{\lower0.3em\hbox{$\,\buildrel >\over\sim\,$}}

\title[Rehabilitating \CIV\ black hole masses]{Rehabilitating \CIV-based black hole mass estimates in quasars}
\author[J. C. Runnoe et al.]{Jessie C. Runnoe$^{1}$\thanks{E-mail:
jrunnoe@uwyo.edu} , M. S. Brotherton$^{1}$, Zhaohui Shang$^{2}$, and M. A. DiPompeo$^1$\\
$^{1}$Department of Physics and Astronomy, University of Wyoming, Laramie, WY 82071, USA\\
$^{2}$Department of Physics, Tianjin Normal University, Tianjin 300387, China}
\begin{document}		

\date{Preprint 2012 May 30}

\pagerange{\pageref{firstpage}--\pageref{lastpage}} \pubyear{2012}

\maketitle

\label{firstpage}

\begin{abstract}
Currently, the ability to produce black hole mass estimates using the \CIVw\ line that are consistent with \Hb\ mass estimates is uncertain, due in large part to disagreement between velocity line width measurements for the two lines.  This discrepancy between \Hb\ and \CIV\ arises from the fact that both line profiles are treated the same way in single-epoch scaling relationships based on the assumption that the broad-line region is virialized, even though a non-virialized emission component is often significant in the \CIV\ line and absent or weak in the \Hb\ line.  Using quasi-simultaneous optical and ultra-violet spectra for a sample of 85 optically bright quasars with redshifts in the range $z=0.03-1.4$, we present a significant step along the path to rehabilitating the \CIV\ line for black hole mass estimates.  We show that the residuals of velocity line width between \CIV\ and \Hb\ are significantly correlated with the peak flux ratio of \siivoivw\ to \CIV.  Using this relationship, we develop a prescription for estimating black hole masses from the ultra-violet spectrum that agree better with \Hb-based masses than the traditional \CIV\ masses.  The scatter between \Hb\ and \CIV\ masses is initially 0.43 dex in our sample and is reduced to 0.33 dex when using our prescription.  The peak flux ratio of \siivoivw\ to \CIV\ is an ultraviolet indicator of the suite of spectral properties commonly known as ``Eigenvector 1'', thus the reduction in scatter between \CIV\ and \Hb\ black hole masses is essentially due to removing an Eigenvector 1 bias in \CIV-based masses.
\end{abstract}


%INTRODUCTION
%%%%%%%%%%%%%%%%%%%%%%%%%%%%%%%%%%%%%%%%%%%%%%%%%%%%%%%%%%%%%%%%%%%%%%%%%%%%%%%%%
\section{introduction}	
%black hole masses are fundamental
Accretion onto a central supermassive black hole is the engine that powers quasars, making the black hole mass a fundamental property of the quasar.  As such, the ability to make accurate measurements of black hole mass is critical for studies of active galactic nuclei (AGN).  There are methods for estimating black hole mass, \citep[see][for a review]{shen13}, but the uncertainties are large and there is room for improvement.

%measure at high-z with VP06 using a line width
In current practice, black hole masses can be estimated for large numbers of AGN with single-epoch spectra using the black hole mass scaling relationships \citep[e.g.,][]{vestergaard06}.  The scaling relationships take measures of velocity of gas in the broad line region (BLR) and distance from the black hole and return a virial mass under the assumption that the BLR motions in the vicinity of the black hole are dominated by its gravity.  The $R-L$ relationship \citep[e.g.,][]{kaspi00,bentz09}, a reverberation mapping result, allows the use of an easily observed continuum luminosity as a proxy for radius and velocity is taken from the line width of Doppler-broadened emission lines.  Commonly used line widths include full-width at half maximum (FWHM) and line dispersion ($\sigma_{l}$), each with their own strengths and weaknesses \citep{denney09a,denney13}.

%issues with CIV
The scaling relationships have been calibrated for multiple emission lines, including \Hb\ and \CIVw\ among others.  The best calibration is for the \Hb\ line because it is the basis for the majority of reverberation mapping programs.  However, at redshifts above $z=1.4$, \Hb\ moves out of the optical wavelength window and the \CIV\ line is more easily obtained.  The reverberation mapping for \CIV\ is more limited but does establish an $R-L$ relationship for \CIV\ which seems to be consistent with the results from \Hb\ reverberation mapping \citep[e.g.,][]{peterson05}.  As a result, there exists a \CIV-based single-epoch mass scaling relationship that should yield mass estimates consistent with those derived from \Hb.

The reliability of the \CIV\ line to reproduce the more trusted \Hb-based black hole mass estimates is not well established.  The main criticism is that the single-epoch \CIV\ profiles do not generally represent the reverberating BLR that provides the basis for virial mass estimates.  This issue manifests itself in several specific issues that limit the successful use of \CIV\ as a virial black hole mass estimator. 

Many studies find very poor agreement and significant scatter between \CIV\ and \Hb\ velocity line widths \citep[e.g.,][]{shen12}.  Reverberation mapping results and virial arguments suggest that the \CIV\ emission originates in a region nearer the central black hole than does the \Hb\ emission \citep{peterson04}.  This would imply that \CIV\ line width measurements should yield broader widths than are measured for \Hb, however the \CIV\ lines widths are often narrower than those of \Hb\ \citep[e.g.][]{baskin05,shang07,trakhtenbrot12}.  Although their velocity line widths are not expected to be equal, if \CIV\ and \Hb\ are both emitted from the BLR they are expected to be correlated.  The fact that they are not suggests that the \CIV\ line width measurements are not probing the velocity of the virialized \CIV\ gas.

Line width measurements may fail to accurately reflect the velocity of the virialized gas because of contamination from other \CIV-emitting regions.  In regular, Type 1 AGN, the \CIV\ line profile is often shifted from the systematic redshift of the object \citep{richards02,baskin05,ho12}.  This implies that the \CIV-emitting gas may not be virialized as there may be a strong wind component to the gas.  However, \citet{vestergaard06} find that this issue is exacerbated by including narrow-line Seyfert 1 sources, low-quality data, absorbed objects, and objects where the \CIV\ narrow-line component is uncertain in the \citet{baskin05} sample, so this issue may be less significant than it appears.  

Reverberation mapping results provide another indication that line width measurements of single-epoch spectra may not reflect the velocity of virialized gas in the BLR.  \citet{denney12} is able to decompose the \CIV\ emission line into a reverberating component and a non-reverberating component by comparing mean spectra, which emulate single-epoch spectra, to rms spectra, which show the emission from gas that responds to changes in the continuum, for objects with reverberation mapping coverage.  The non-reverberating component contributes a fairly narrow core to the \CIV\ line that draws the FWHM to lower values.  A measure of line width in a single-epoch spectrum that might be used in a black hole mass scaling relationship is unable to separate this component from the virialized emission.  \citet{denney09a} has simulated the effects of including a non-virialized narrow component in the line widths used for black hole mass calculations.  The results are conclusive: to obtain accurate and precise line width estimates, particularly when using FWHM, such contamination must be removed.

The narrow core component is much less significant in \Hb\ than in \CIV, introducing scatter between \Hb\ and \CIV-based black hole mass estimates.  \citet{denney12} recommends the shape parameter, $S=\rm{FWHM}/\sigma_{l}$, for distinguishing between objects with boxy profiles ($S\sim1.5$), indicative of a strong reverberating component, and peaky ($S\sim0.5$) profiles, indicative of a strong non-reverberating component.  Furthermore, she provides a prescription for using the shape parameter to remove the bias in \CIV\ black hole mass estimates.  The shape parameter depends on \CIV\ FWHM, and is a clear source of scatter between \Hb\ and \CIV\ black hole masses.

%EV1
The shape parameter, which can be used to isolate the component of the \CIV\ emission line that originates in virialized gas and improve the reliability of \CIV\ black hole mass estimates, is actually one among many correlated spectral properties collectively termed ``Eigenvector 1'' (EV1).  EV1 describes the largest amount of variation between quasar spectra \citep{bg92}, and is dominated by the anti-correlation of optical \FeII\ and \OIIIw\ emission, with objects on one end having strong \FeII\ emission and weak \OIII\ emission.  Originally defined at rest-frame optical wavelengths, EV1 has ultra-violet (UV) indicators \citep{brotherton99a,shang03,sulentic07}.  There have been suggestions for physical drivers of EV1, including the Eddington ratio at low redshifts \citep{bg92}, but it is not clear if this is the case and whether it holds at higher redshifts \citep{yuan03}.  Although the exact relationship between EV1 and black hole mass remains elusive, black hole mass is not the primary driver of EV1.  Objects with similar black hole masses exhibit a range of EV1 properties as demonstrated in figure 6 of \citet{boroson02} and therefore, crucially for this work, an object's location on EV1 is not indicative of its intrinsic black hole mass.  In some samples (e.g., the Palomar-Green sample) EV1 and black hole mass appear to be correlated but this may be the result of the sample selection and other primary correlations like the one between EV1 and Eddington fraction.

%more robust measures of EV1
The spectral analysis of \citet{wills93b} suggests that it is in fact an EV1 bias that causes disagreement between the \CIV\ and \Hb\ velocity widths.  Their UV spectral composites demonstrate that with increasing \CIV\ FWHM, the peak flux of the \CIV\ line decreases compared to the peak flux of the nearby \siivoiv\ blend at 1400 \AA\ \citep[which does not vary with the FWHM of \CIV,][]{wills93b,richards02} and the line profile profile changes from peaky to boxy.  As a result, the velocity width of the \CIV\ emission line is determined both by the the black hole mass and the unknown physical driver of EV1.  The shape of \CIV\ is a UV indicator of EV1, suggesting that the shape correction of \citet{denney12} is successful largely because of its ability to estimate the EV1 bias in \CIV.  However, line dispersion, and thus the shape parameter, requires high signal-to-noise spectra to obtain an accurate measurement and is thus not suitable for use with current survey-quality data \citep{denney13}.  Instead, the application of a  peak ratio of \siivoiv\ to \CIV, a UV EV1 indicator that is free of the S/N constraint that plagues the shape parameter, combined with \CIV\ FWHM-based black hole mass estimates might provide a more \Hb-like black hole mass that can be measured for large numbers of quasars.  

%we can correct this
%to dramatically reduce the scatter in CIV Mbh
We investigate the use of the ratio of the \siivoiv\ to \CIV\ peak fluxes as a method to rehabilitate the \CIV-based black hole mass estimates in the spectral energy distribution (SED) atlas of \citet{shang11}.  The quasi-simultaneous optical and UV spectrophotometry available in this data set is particularly valuable to this investigation, allowing us to consider the \Hb\ and \CIV\ spectral regions free of variability concerns.  We show that by using both the line width of \CIV\ and peak ratio it is possible to dramatically reduce the scatter between \Hb\ and \CIV-based black hole mass estimates and derive prescriptions for predicting the \Hb\ line width and black hole mass from information in the \CIV\ spectral region.    

This investigation is organized as follows.  In Section~\ref{sec:data} we present the sample and discuss the our methods for making spectral measurements.  Section~\ref{sec:analysis} presents an analysis of the dependence of velocity line width residuals on the ratio of peak fluxes of \siivoiv\ to \CIV, including a correction for the observed effect that decreases the scatter between \CIV\ and \Hb-based black hole masses.  The results are discussed in the context of other work in Section~\ref{sec:discussion} and conclusions are presented in Section~\ref{sec:conclusion}.  Throughout this work back hole masses and luminosities are calculated using a cosmology with $H_0 = 70$ km s$^{-1}$ Mpc$^{-1}$, $\Omega_{\Lambda} = 0.7$, and $\Omega_{m} = 0.3$.

%DATA
%%%%%%%%%%%%%%%%%%%%%%%%%%%%%%%%%%%%%%%%%%%%%%%%%%%%%%%%%%%%%%%%%%%%%%%%%%%%%%%%%
\section{Sample, Data, and Measurements}
\label{sec:data}
%sample
For this work we use the SED sample from \citet{shang11}.  The atlas includes 85 objects, 58 of which are radio-loud (RL) and 27 of which are radio-quiet (RQ), with redshifts in the range $z=0.030-1.404$ and bolometric luminosities of $45.1<\textrm{log}(L_{bol})<47.3$ in units of ergs per second.	

Most objects have quasi-simultaneous optical and UV spectrophotometry.  All of the objects were observed in the UV either with the {\it Hubble Space Telescope} Faint Object Spectrograph (FOS) or Space Telescope Imaging Spectrograph (STIS) and most were followed up within a few weeks by low-resolution ground-base optical spectrophotometry from McDonald Observatory or the Kitt Peak National Observatory. 

The SED atlas does not comprise a complete sample, but it does include a nearly complete subsample.  The PGX subsample has 22 out of 23 Palomar-Green (PG) quasars selected to study the soft X-ray regime \citep{laor94,laor97}, is UV bright, and has  $z \le 0.4$.  This subsample was presented in \citet{shang07}, who found that the line widths for these objects do not indicate a purely stratified ionization structure in the BLR.  Furthermore, these objects have also been used to illustrate the issues associated with \CIV\ line widths \citep[e.g.,][]{trakhtenbrot12}.  The atlas also includes the higher-redshift, higher-luminosity RL subsample and the FUSE subsample that contains mostly radio-quiet objects with redshifts of a few tenths.

This atlas is particularly suited to this investigation for two reasons.  First, the quasi-simultaneous optical and UV spectrophotomety makes this sample less susceptible to variability issues than similar investigations \citep[e.g.,][]{trakhtenbrot12}, allowing a meaningful comparison between physical properties calculated from the \CIV\ and \Hb\ wavelength regimes.  Second, the inclusion of the PGX subsample ensures that the total atlas covers the full range in EV1 properties.  This is important because the sample contains a large number of RL sources which are isolated on one end of EV1 with strong \OIII\ and weak \FeII\ \citep[e.g.,][]{bg92,bachev04,sulentic07}.  The sample will exhibit EV1 properties within the range defined by the PGX subsample making our conclusions robust, the fact that the overabundance of RL sources may cause the distribution of those properties to be biased does not compromise this.	
	
The spectra have been corrected for Galactic extinction, reddening, and host contamination in the optical (although these objects are optically bright and the host contribution is minimal).  The redshift is determined from the \OIIIw\ line and the spectra were shifted to the rest frame.  \citet{shang11} provides more details on these corrections. 

%fitting
Spectral fitting was done by \citet{tang12} following the method of \citet{shang07}.  Fitting is done by minimizing $\chi^{2}$ between the model and the data using the \textsc{IRAF} package \textsc{SPECFIT} \citep{kriss94}.  The \Hb, \CIVw, and \siivoivw\ regions were fit individually with model spectra consisting of a power-law continuum and two Gaussian components per broad emission line.  In the case of the \siivoiv\ blend, which we will hereafter refer to as ``$\lambda1400$'', the fit was made between 1350 and 1460 \AA\ and the power-law continuum is separate from that of \CIV.  The Gaussians ascribed to \SiIV\ and \OIV\ were identical except in their centroids which were appropriate for their respective line centers.  For the \CIV\ doublet, the Gaussian pairs were also identical except in their centroids and they did not include a component to model emission from the narrow-line region (NLR) because, according to \citet{wills93}, there is no strong NLR component present in \CIV.  The \Hb\ broad emission line does include a contribution to emission from the NLR which was modeled with an additional Gaussian that is not included when measuring line widths.  A \FeII\ template derived from the narrow-line Seyfert 1 I Zw1 \citep{bg92} was also included in the \Hb\ region and was allowed to vary in amplitude and velocity width in order to match the observed spectrum.

\begin{figure}
\begin{center}
\includegraphics[width=8.9 truecm]{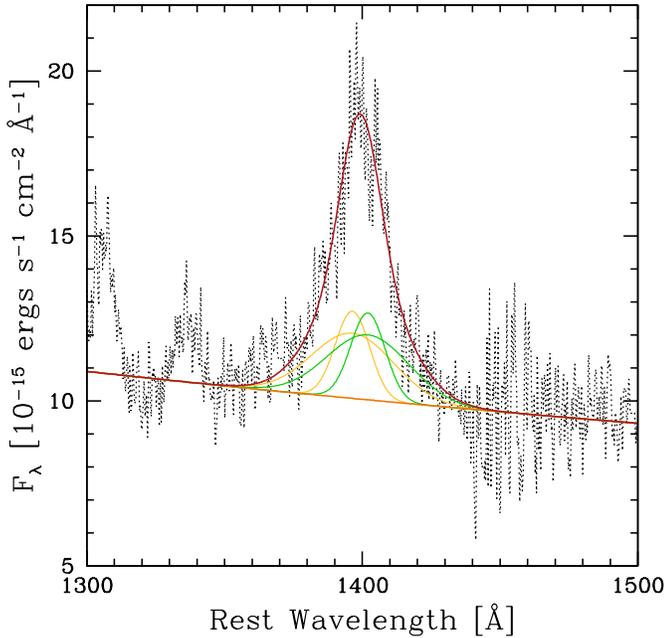}
\end{center}
\caption{Spectral fitting for the wavelength region around $\lambda1400$ for PG 1415+451.  The observed spectrum is shown by the dotted gray line, the total model spectrum by the red line, the power-law continuum by the orange line, the two \SiIV\ components by the gold lines, and the two \OIV\ components by the green lines.}
\label{fig:spex}
\end{figure}

The spectral fits for $\lambda1400$ were done at the same time as the \Hb\ and \CIV\ regions but not presented in \citet{tang12} so we present them here.  A sample spectrum is given in Figure~\ref{fig:spex} and the results of the fitting are given in Tables~\ref{tab:EV1} and \ref{tab:phys}.

%errors
In most cases, errors on the fit parameters were calculated by \citet{tang12} \citep[see also,][]{shang07,laor94a}.  For spectra with high S/N, the largest uncertainty comes from placing the continuum in each wavelength region rather than formal errors from the fitting process.  Therefore, they estimated the errors in the continuum placement by calculating the 1$\sigma$ errors on the flux in the observed spectrum at each end of the local continuum.  They then calculate four new continua, corresponding to adding and subtracting the 1$\sigma$ flux errors on each end of the local continuum, and recalculate the measured quantities.  The largest positive and negative changes in the measured quantities were taken to be the errors.  Uncertainty in the EW and flux in the line measurements for the \Hb\ and \CIV\ regions that are not listed here are given in \citet{tang12} and calculated via the above method.

The above method for calculating uncertainties is problematic for some parameters (namely line dispersion), which led us to perform a Monte-Carlo fitting procedure similar to that of \citet{dipompeo12b} to determine the uncertainties on these parameters.  The resulting errors on all fit parameters were very small, indicating that the formal errors on our fitting procedure are small.  The uncertainties on each parameter are actually a combination of the fit uncertainties and systematic uncertainties, where the systematic uncertainties likely dominate in our sample.

For the FWHM measurements, which are at the core of this analysis, it is particularly important not to underestimate the measurement uncertainties by ignoring systematic uncertainties.  A comparison of measurements  of the FWHM of \CIV\ made with similar but not identical methodologies in the literature \citep{vestergaard06,shang07} for some objects in our sample shows that measurements typically vary by 15\%.  This is larger than all but two of the percent errors listed for \CIV\ FWHM (the largest of which is 16\%) in Table~\ref{tab:phys}, suggesting that the errors on FWHM may be underestimated.  We expect that uncertainty in the line dispersion parameter will be at least as large as for FWHM, indicating that these are likely also at least 15\%.  While applying our fitting procedure across the literature might yield more consistent results than this, we must also consider the possibility that different fitting procedures behave systematically differently for different line profiles.  Thus, in order to be appropriately conservative, we adopt 15\% uncertainty for all velocity line width measurements.  

Measurements associated with the optical and UV indicators of EV1 are provided in Table~\ref{tab:EV1}.  \citet{runnoe12a} measured continuum luminosities for this sample in the \Hb\ and \CIV\ wavelength regimes that are listed in Table~\ref{tab:phys} along with line width measurements.

\begin{table*}
\begin{minipage}{18cm}
\caption{Eigenvector 1 measurements}
\label{tab:EV1}
\scalebox{0.7}{
\begin{tabular}{lcccccccc}
Object  & $\lambda1400$ EW & $\lambda1400$ Peak & $\lambda1400$ Flux & \CIV\ EW & \CIV\ Peak & \CIV\ Flux & \FeII\ EW  & \OIII\ EW \\
  & \AA\ \footnotetext{Note $-$ EWs are given in the rest frame and fluxes in the observed frame.} & $10^{-15}$ ergs s$^{-1}$ cm$^{-2}$ \AA$^{-1}$ & $10^{-15}$ ergs s$^{-1}$ cm$^{-2}$ & \AA & $10^{-15}$ ergs s$^{-1}$ cm$^{-2}$ \AA$^{-1}$ & $10^{-15}$ ergs s$^{-1}$ cm$^{-2}$ & \AA  & \AA \\
\hline
3C 110   
	& 
   12.77
$\pm$    0.26
 & 
    0.75
$\pm$    0.02
 & 
   50.87
$\pm$    0.98
 & 
  107.78
 & 
    7.84
$\pm$    0.02
 & 
  363.39
 & 
   27.08
 & 
   24.98
 \\
3C 175   
	& 
   12.56
$\pm$    0.59
 & 
    0.43
$\pm$    0.02
 & 
   37.53
$\pm$    0.95
 & 
   55.59
 & 
    3.24
$\pm$    0.02
 & 
  158.61
 & 
   21.10
 & 
   20.05
 \\
3C 186   
	& 
   15.70
$\pm$    1.20
 & 
    0.15
$\pm$    0.02
 & 
   10.52
$\pm$    0.93
 & 
   66.50
 & 
    0.91
$\pm$    0.02
 & 
   40.18
 & 
\nodata
 & 
\nodata
 \\
3C 207   
	& 
   15.21
$\pm$    0.88
 & 
    0.38
$\pm$    0.02
 & 
   12.99
$\pm$    0.94
 & 
   86.32
 & 
    2.31
$\pm$    0.02
 & 
   75.68
 & 
   60.97
 & 
   23.34
 \\
3C 215   
	& 
   18.88
$\pm$    0.56
 & 
    0.53
$\pm$    0.02
 & 
   15.49
$\pm$    0.95
 & 
  204.60
 & 
    3.65
$\pm$    0.02
 & 
  153.49
 & 
   55.13
 & 
   47.03
 \\
3C 232   
	& 
   14.06
$\pm$    0.24
 & 
    1.11
$\pm$    0.02
 & 
   45.03
$\pm$    0.98
 & 
   33.54
 & 
    2.30
$\pm$    0.03
 & 
  100.00
 & 
   66.87
 & 
   55.39
 \\
3C 254   
	& 
   19.84
$\pm$    1.11
 & 
    0.33
$\pm$    0.02
 & 
   13.33
$\pm$    0.93
 & 
  172.43
 & 
    2.44
$\pm$    0.02
 & 
  107.46
 & 
  102.79
 & 
  116.51
 \\
3C 263   
	& 
   12.52
$\pm$    0.36
 & 
    2.01
$\pm$    0.02
 & 
   68.07
$\pm$    0.97
 & 
   75.51
 & 
   11.50
$\pm$    0.02
 & 
  344.69
 & 
   58.81
 & 
   20.10
 \\
3C 277.1   
	& 
   17.56
$\pm$    0.88
 & 
    1.07
$\pm$    0.02
 & 
   30.54
$\pm$    0.94
 & 
  106.47
 & 
    6.42
$\pm$    0.02
 & 
  163.87
 & 
  103.86
 & 
  159.40
 \\
3C 281   
	& 
   15.83
$\pm$    0.97
 & 
    0.38
$\pm$    0.02
 & 
   22.56
$\pm$    0.93
 & 
  119.67
 & 
    3.45
$\pm$    0.02
 & 
  149.15
 & 
   62.32
 & 
   46.79
 \\
3C 288.1   
	& 
   10.17
$\pm$    0.41
 & 
    0.13
$\pm$    0.02
 & 
    7.97
$\pm$    0.96
 & 
   42.32
 & 
    0.82
$\pm$    0.02
 & 
   29.92
 & 
  121.72
 & 
   85.38
 \\
3C 334   
	& 
   14.67
$\pm$    0.23
 & 
    0.84
$\pm$    0.02
 & 
   53.29
$\pm$    0.98
 & 
   74.98
 & 
    6.35
$\pm$    0.02
 & 
  250.62
 & 
   26.22
 & 
   40.53
 \\
3C 37   
	& 
   34.00
$\pm$    2.54
 & 
    0.47
$\pm$    0.01
 & 
   14.54
$\pm$    0.90
 & 
  252.63
 & 
    3.64
$\pm$    0.02
 & 
   97.43
 & 
  204.08
 & 
  105.93
 \\
3C 446   
	& 
   18.16
$\pm$    6.35
 & 
    0.10
$\pm$    0.01
 & 
    6.07
$\pm$    0.87
 & 
   76.40
 & 
    0.95
$\pm$    0.02
 & 
   28.00
 & 
\nodata
 & 
\nodata
 \\
3C 47   
	& 
   18.30
$\pm$    1.14
 & 
    0.79
$\pm$    0.02
 & 
   23.17
$\pm$    0.93
 & 
  172.91
 & 
    4.98
$\pm$    0.02
 & 
  212.37
 & 
   67.08
 & 
  116.19
 \\
4C 01.04   
	& 
   50.12
$\pm$    4.21
 & 
    0.46
$\pm$    0.01
 & 
   28.56
$\pm$    0.88
 & 
  270.36
 & 
    3.82
$\pm$    0.02
 & 
  173.77
 & 
   95.02
 & 
   39.04
 \\
4C 06.69   
	& 
   10.52
$\pm$    0.64
 & 
    0.70
$\pm$    0.02
 & 
   38.24
$\pm$    0.95
 & 
   45.16
 & 
    3.71
$\pm$    0.02
 & 
  150.95
 & 
   42.11
 & 
   42.89
 \\
4C 10.06   
	& 
   21.33
$\pm$    1.31
 & 
    2.58
$\pm$    0.02
 & 
  121.57
$\pm$    0.92
 & 
  114.57
 & 
   17.44
$\pm$    0.02
 & 
  598.92
 & 
   59.90
 & 
   24.80
 \\
4C 11.69   
	& 
   11.98
$\pm$    0.33
 & 
    0.38
$\pm$    0.02
 & 
   23.02
$\pm$    0.97
 & 
   37.41
 & 
    2.51
$\pm$    0.02
 & 
   65.65
 & 
\nodata
 & 
\nodata
 \\
4C 12.40   
	& 
   22.88
$\pm$    1.05
 & 
    0.24
$\pm$    0.02
 & 
   14.50
$\pm$    0.93
 & 
   97.16
 & 
    1.51
$\pm$    0.02
 & 
   59.50
 & 
  142.53
 & 
   40.87
 \\
4C 19.44   
	& 
   11.90
$\pm$    0.37
 & 
    1.52
$\pm$    0.02
 & 
   44.04
$\pm$    0.96
 & 
   96.37
 & 
   11.94
$\pm$    0.02
 & 
  309.71
 & 
   69.49
 & 
   78.62
 \\
4C 20.24   
	& 
   25.06
$\pm$    1.03
 & 
    0.78
$\pm$    0.02
 & 
   23.20
$\pm$    0.93
 & 
  185.25
 & 
    4.76
$\pm$    0.02
 & 
  136.93
 & 
\nodata
 & 
\nodata
 \\
4C 22.26   
	& 
   16.96
$\pm$    1.26
 & 
    0.22
$\pm$    0.02
 & 
    8.78
$\pm$    0.92
 & 
  180.90
 & 
    2.00
$\pm$    0.02
 & 
   78.42
 & 
\nodata
 & 
\nodata
 \\
4C 30.25   
	& 
   17.76
$\pm$    1.24
 & 
    0.14
$\pm$    0.02
 & 
    4.56
$\pm$    0.92
 & 
  146.34
 & 
    1.09
$\pm$    0.02
 & 
   32.92
 & 
\nodata
 & 
\nodata
 \\
4C 31.63   
	& 
    8.31
$\pm$    0.43
 & 
    7.48
$\pm$    0.02
 & 
  254.54
$\pm$    0.96
 & 
   39.60
 & 
   29.13
$\pm$    0.02
 & 
 1109.30
 & 
  114.45
 & 
    6.44
 \\
4C 34.47   
	& 
   23.15
$\pm$    1.46
 & 
    7.47
$\pm$    0.02
 & 
  201.50
$\pm$    0.92
 & 
  213.03
 & 
   67.11
$\pm$    0.02
 & 
 1802.20
 & 
  167.16
 & 
   92.68
 \\
4C 39.25   
	& 
    9.75
$\pm$    0.13
 & 
    0.68
$\pm$    0.02
 & 
   37.76
$\pm$    1.00
 & 
   79.99
 & 
    6.75
$\pm$    0.03
 & 
  257.07
 & 
   21.95
 & 
   15.20
 \\
4C 40.24   
	& 
   15.36
$\pm$    1.16
 & 
    0.22
$\pm$    0.02
 & 
    5.36
$\pm$    0.93
 & 
  131.38
 & 
    1.21
$\pm$    0.02
 & 
   38.00
 & 
\nodata
 & 
\nodata
 \\
4C 41.21   
	& 
   18.88
$\pm$    0.38
 & 
    2.46
$\pm$    0.02
 & 
   99.06
$\pm$    0.96
 & 
   96.04
 & 
   14.76
$\pm$    0.02
 & 
  451.29
 & 
   61.78
 & 
   32.98
 \\
4C 49.22   
	& 
   24.05
$\pm$    1.25
 & 
    1.77
$\pm$    0.02
 & 
   62.16
$\pm$    0.92
 & 
  178.77
 & 
   11.80
$\pm$    0.02
 & 
  379.18
 & 
  156.32
 & 
   26.63
 \\
4C 55.17   
	& 
    4.32
$\pm$    0.60
 & 
    0.05
$\pm$    0.02
 & 
    2.65
$\pm$    0.95
 & 
   34.57
 & 
    0.53
$\pm$    0.02
 & 
   20.00
 & 
\nodata
 & 
\nodata
 \\
4C 58.29   
	& 
    5.59
$\pm$    0.79
 & 
    0.13
$\pm$    0.02
 & 
    8.03
$\pm$    1.07
 & 
   39.43
 & 
    1.20
$\pm$    0.02
 & 
   49.09
 & 
\nodata
 & 
\nodata
 \\
4C 64.15   
	& 
   17.89
$\pm$    1.55
 & 
    0.11
$\pm$    0.02
 & 
    6.88
$\pm$    0.92
 & 
   68.54
 & 
    0.52
$\pm$    0.02
 & 
   26.00
 & 
\nodata
 & 
\nodata
 \\
4C 73.18   
	& 
   11.58
$\pm$    0.63
 & 
    6.02
$\pm$    0.02
 & 
  158.98
$\pm$    0.95
 & 
  111.79
 & 
   38.03
$\pm$    0.02
 & 
 1119.60
 & 
   41.19
 & 
   24.41
 \\
B2 0742+31   
	& 
   13.29
$\pm$    0.36
 & 
    1.84
$\pm$    0.02
 & 
   69.12
$\pm$    0.97
 & 
  127.78
 & 
   13.99
$\pm$    0.02
 & 
  550.29
 & 
   31.21
 & 
   41.31
 \\
B2 1351+31   
	& 
    7.66
$\pm$    0.83
 & 
    0.05
$\pm$    0.02
 & 
    2.89
$\pm$    0.94
 & 
   47.73
 & 
    0.39
$\pm$    0.02
 & 
   14.73
 & 
\nodata
 & 
\nodata
 \\
B2 1555+33   
	& 
   18.84
$\pm$    2.28
 & 
    0.13
$\pm$    0.01
 & 
    5.37
$\pm$    0.90
 & 
  108.54
 & 
    0.91
$\pm$    0.02
 & 
   30.00
 & 
\nodata
 & 
\nodata
 \\
B2 1611+34   
	& 
   13.03
$\pm$    1.06
 & 
    0.24
$\pm$    0.01
 & 
   10.21
$\pm$    0.74
 & 
   50.65
 & 
    1.03
$\pm$    0.01
 & 
   35.85
 & 
    0.00
 & 
   17.44
 \\
IRAS F07546+3928   
	& 
    7.83
$\pm$    0.69
 & 
    6.55
$\pm$    0.02
 & 
  117.47
$\pm$    0.94
 & 
  105.34
 & 
   70.56
$\pm$    0.02
 & 
 1640.60
 & 
  188.68
 & 
   38.01
 \\
MC2 0042+101   
	& 
   28.15
$\pm$    4.01
 & 
    0.17
$\pm$    0.01
 & 
    7.82
$\pm$    0.90
 & 
  218.94
 & 
    1.65
$\pm$    0.02
 & 
   61.92
 & 
    0.00
 & 
   50.37
 \\
MC2 1146+111   
	& 
\nodata

 & 
\nodata

 & 
\nodata

 & 
   46.76
 & 
    0.67
$\pm$    0.02
 & 
   21.04
 & 
  111.28
 & 
   22.07
 \\
MRK 506   
	& 
   38.46
$\pm$    1.28
 & 
   12.91
$\pm$    0.02
 & 
  454.64
$\pm$    0.92
 & 
  244.42
 & 
   75.10
$\pm$    0.02
 & 
 2670.80
 & 
  101.49
 & 
   35.72
 \\
MRK 509   
	& 
   24.41
$\pm$    0.09
 & 
   63.26
$\pm$    0.02
 & 
 1988.98
$\pm$    1.01
 & 
  140.75
 & 
  316.41
$\pm$    0.17
 & 
10404.00
 & 
   99.66
 & 
   83.64
 \\
OS 562   
	& 
   11.70
$\pm$    0.54
 & 
    0.66
$\pm$    0.02
 & 
   30.01
$\pm$    0.95
 & 
   46.38
 & 
    3.50
$\pm$    0.02
 & 
  105.04
 & 
   77.36
 & 
   14.43
 \\
PG 0052+251   
	& 
   15.65
$\pm$    1.00
 & 
    7.98
$\pm$    0.02
 & 
  284.58
$\pm$    0.93
 & 
  137.58
 & 
   51.26
$\pm$    0.02
 & 
 2181.00
 & 
   52.62
 & 
   53.81
 \\
PG 0844+349   
	& 
   15.12
$\pm$    0.21
 & 
   16.29
$\pm$    0.02
 & 
  455.52
$\pm$    0.99
 & 
   48.19
 & 
   46.03
$\pm$    0.03
 & 
 1332.70
 & 
  211.82
 & 
    7.13
 \\
PG 0947+396   
	& 
   11.87
$\pm$    0.72
 & 
    3.01
$\pm$    0.02
 & 
   81.63
$\pm$    0.94
 & 
   95.66
 & 
   18.56
$\pm$    0.02
 & 
  590.26
 & 
  129.44
 & 
   23.68
 \\
PG 0953+414   
	& 
    6.67
$\pm$    0.52
 & 
    7.74
$\pm$    0.02
 & 
  178.25
$\pm$    0.95
 & 
   49.35
 & 
   42.75
$\pm$    0.02
 & 
 1150.10
 & 
   42.99
 & 
   12.04
 \\
PG 1001+054   
	& 
   26.76
$\pm$    0.83
 & 
    5.23
$\pm$    0.02
 & 
  104.68
$\pm$    0.94
 & 
   75.08
 & 
   10.01
$\pm$    0.02
 & 
  272.12
 & 
  202.69
 & 
   12.11
 \\
PG 1100+772   
	& 
    4.32
$\pm$    0.83
 & 
    3.29
$\pm$    0.02
 & 
   63.84
$\pm$    0.94
 & 
   79.68
 & 
   22.99
$\pm$    0.02
 & 
  950.62
 & 
   54.83
 & 
   42.10
 \\
PG 1103-006   
	& 
   14.56
$\pm$    0.85
 & 
    1.54
$\pm$    0.02
 & 
   51.79
$\pm$    0.94
 & 
   55.29
 & 
    5.18
$\pm$    0.02
 & 
  181.26
 & 
   97.99
 & 
   13.08
 \\
PG 1114+445   
	& 
   17.48
$\pm$    1.71
 & 
    3.04
$\pm$    0.02
 & 
  110.01
$\pm$    0.91
 & 
   81.06
 & 
   15.27
$\pm$    0.02
 & 
  497.82
 & 
   48.70
 & 
   15.75
 \\
PG 1115+407   
	& 
   14.19
$\pm$    0.76
 & 
    6.50
$\pm$    0.02
 & 
  181.79
$\pm$    0.94
 & 
   47.59
 & 
   14.90
$\pm$    0.02
 & 
  519.46
 & 
  149.95
 & 
    8.05
 \\
PG 1116+215   
	& 
   15.02
$\pm$    0.47
 & 
   16.27
$\pm$    0.02
 & 
  523.42
$\pm$    0.96
 & 
   71.84
 & 
   70.19
$\pm$    0.02
 & 
 2160.80
 & 
  171.68
 & 
   16.72
 \\
PG 1202+281   
	& 
   39.63
$\pm$    1.95
 & 
    3.66
$\pm$    0.01
 & 
   98.11
$\pm$    0.91
 & 
  306.34
 & 
   28.91
$\pm$    0.02
 & 
  711.04
 & 
   59.75
 & 
   55.91
 \\
PG 1216+069   
	& 
   10.27
$\pm$    0.24
 & 
    1.99
$\pm$    0.02
 & 
   68.10
$\pm$    0.98
 & 
   98.03
 & 
   17.29
$\pm$    0.03
 & 
  557.16
 & 
   41.99
 & 
   10.49
 \\
PG 1226+023   
	& 
    5.91
$\pm$    0.52
 & 
   55.37
$\pm$    0.02
 & 
 1447.16
$\pm$    0.95
 & 
   32.40
 & 
  222.96
$\pm$    0.02
 & 
 7417.20
 & 
  111.46
 & 
    8.30
 \\
PG 1259+593   
	& 
   11.29
$\pm$    0.40
 & 
    3.21
$\pm$    0.02
 & 
  112.42
$\pm$    0.96
 & 
   20.31
 & 
    4.33
$\pm$    0.02
 & 
  184.00
 & 
  403.05
 & 
    4.45
 \\
PG 1309+355   
	& 
    8.25
$\pm$    0.58
 & 
    3.89
$\pm$    0.02
 & 
   91.00
$\pm$    0.95
 & 
   56.31
 & 
   16.51
$\pm$    0.02
 & 
  549.79
 & 
   66.20
 & 
   17.05
 \\
PG 1322+659   
	& 
    7.53
$\pm$    1.46
 & 
    2.42
$\pm$    0.02
 & 
   72.00
$\pm$    0.92
 & 
   54.44
 & 
   15.67
$\pm$    0.02
 & 
  454.92
 & 
   97.48
 & 
    7.49
 \\
PG 1351+640   
	& 
   22.06
$\pm$    0.52
 & 
   17.53
$\pm$    0.02
 & 
  358.64
$\pm$    0.95
 & 
   78.63
 & 
   49.34
$\pm$    0.02
 & 
 1296.40
 & 
   29.26
 & 
   38.10
 \\
PG 1352+183   
	& 
   17.48
$\pm$    1.22
 & 
    5.13
$\pm$    0.02
 & 
  174.24
$\pm$    0.92
 & 
   77.75
 & 
   20.50
$\pm$    0.02
 & 
  650.23
 & 
  112.52
 & 
    9.68
 \\
PG 1402+261   
	& 
   14.05
$\pm$    0.82
 & 
    9.44
$\pm$    0.02
 & 
  301.14
$\pm$    0.94
 & 
   44.60
 & 
   24.73
$\pm$    0.02
 & 
  831.74
 & 
  259.46
 & 
    2.45
 \\
PG 1411+442   
	& 
   17.40
$\pm$    0.40
 & 
   20.56
$\pm$    0.02
 & 
  426.44
$\pm$    0.96
 & 
   44.69
 & 
   63.08
$\pm$    0.02
 & 
  911.68
 & 
  122.90
 & 
   13.37
 \\
PG 1415+451   
	& 
   22.81
$\pm$    2.46
 & 
    8.63
$\pm$    0.01
 & 
  229.08
$\pm$    0.90
 & 
   65.49
 & 
   23.30
$\pm$    0.02
 & 
  588.64
 & 
  210.30
 & 
    2.68
 \\
PG 1425+267   
	& 
   10.30
$\pm$    0.50
 & 
    1.25
$\pm$    0.02
 & 
   41.86
$\pm$    0.95
 & 
  123.53
 & 
    9.07
$\pm$    0.02
 & 
  449.43
 & 
   34.43
 & 
   31.57
 \\
PG 1427+480   
	& 
   16.11
$\pm$    0.64
 & 
    3.47
$\pm$    0.02
 & 
  108.83
$\pm$    0.95
 & 
   77.59
 & 
   21.38
$\pm$    0.02
 & 
  463.34
 & 
   83.29
 & 
   27.86
 \\
PG 1440+356   
	& 
   11.62
$\pm$    0.47
 & 
   23.43
$\pm$    0.02
 & 
  489.35
$\pm$    0.96
 & 
   33.02
 & 
   78.97
$\pm$    0.02
 & 
 1246.80
 & 
  230.09
 & 
    7.59
 \\
PG 1444+407   
	& 
   14.67
$\pm$    0.29
 & 
    3.91
$\pm$    0.02
 & 
  132.63
$\pm$    0.97
 & 
   24.09
 & 
    6.70
$\pm$    0.02
 & 
  198.97
 & 
  287.49
 & 
    1.49
 \\
PG 1512+370   
	& 
    6.95
$\pm$    0.27
 & 
    0.93
$\pm$    0.02
 & 
   28.00
$\pm$    0.98
 & 
  119.28
 & 
   10.43
$\pm$    0.02
 & 
  420.14
 & 
   23.43
 & 
   58.42
 \\
PG 1534+580   
	& 
   16.05
$\pm$    1.74
 & 
    7.18
$\pm$    0.02
 & 
  213.60
$\pm$    0.91
 & 
  142.51
 & 
   65.41
$\pm$    0.02
 & 
 1740.30
 & 
   25.40
 & 
  114.48
 \\
PG 1543+489   
	& 
   18.20
$\pm$    0.22
 & 
    3.19
$\pm$    0.02
 & 
  100.46
$\pm$    0.98
 & 
   39.07
 & 
    5.69
$\pm$    0.03
 & 
  209.09
 & 
  206.92
 & 
    5.42
 \\
PG 1545+210   
	& 
   20.56
$\pm$    0.98
 & 
    3.58
$\pm$    0.02
 & 
  143.52
$\pm$    0.93
 & 
  181.05
 & 
   26.77
$\pm$    0.02
 & 
 1047.40
 & 
   24.95
 & 
   41.54
 \\
PG 1626+554   
	& 
   16.67
$\pm$    1.26
 & 
    7.21
$\pm$    0.02
 & 
  281.87
$\pm$    0.92
 & 
   70.85
 & 
   30.96
$\pm$    0.02
 & 
 1012.60
 & 
   72.18
 & 
    5.44
 \\
PG 1704+608   
	& 
    2.68
$\pm$    0.38
 & 
    0.42
$\pm$    0.02
 & 
   23.79
$\pm$    0.96
 & 
   61.20
 & 
   11.99
$\pm$    0.02
 & 
  495.60
 & 
   13.60
 & 
   27.68
 \\
PG 2214+139   
	& 
   12.94
$\pm$    0.13
 & 
   15.89
$\pm$    0.02
 & 
  562.97
$\pm$    0.99
 & 
   51.88
 & 
   76.43
$\pm$    0.02
 & 
 2100.90
 & 
  173.28
 & 
   12.43
 \\
PG 2251+113   
	& 
   12.12
$\pm$    0.39
 & 
    3.33
$\pm$    0.02
 & 
  103.58
$\pm$    0.96
 & 
   96.91
 & 
   19.56
$\pm$    0.02
 & 
  713.28
 & 
  144.98
 & 
   28.07
 \\
PG 2349-014   
	& 
   10.58
$\pm$    1.82
 & 
    3.53
$\pm$    0.01
 & 
   93.80
$\pm$    0.91
 & 
  155.39
 & 
   30.32
$\pm$    0.02
 & 
 1231.60
 & 
  106.56
 & 
   32.27
 \\
PKS 0112-017   
	& 
   13.00
$\pm$    0.28
 & 
    0.19
$\pm$    0.02
 & 
    8.98
$\pm$    0.97
 & 
   30.81
 & 
    0.54
$\pm$    0.03
 & 
   20.00
 & 
\nodata
 & 
\nodata
 \\
PKS 0403-13   
	& 
   15.60
$\pm$    0.62
 & 
    0.97
$\pm$    0.02
 & 
   29.98
$\pm$    1.05
 & 
  138.32
 & 
    7.92
$\pm$    0.02
 & 
  213.37
 & 
   67.08
 & 
   19.87
 \\
PKS 0859-14   
	& 
   10.09
$\pm$    0.62
 & 
    0.46
$\pm$    0.02
 & 
   14.73
$\pm$    0.95
 & 
   48.89
 & 
    1.80
$\pm$    0.02
 & 
   59.88
 & 
   94.14
 & 
    8.67
 \\
PKS 1127-14   
	& 
   12.62
$\pm$    2.13
 & 
    0.25
$\pm$    0.01
 & 
   15.00
$\pm$    0.91
 & 
   25.93
 & 
    1.03
$\pm$    0.02
 & 
   30.00
 & 
\nodata
 & 
\nodata
 \\
PKS 1656+053   
	& 
    6.35
$\pm$    0.66
 & 
    0.40
$\pm$    0.02
 & 
   17.06
$\pm$    0.95
 & 
\nodata
 & 
\nodata

 & 
\nodata
 & 
  114.08
 & 
    3.84
 \\
PKS 2216-03   
	& 
   15.36
$\pm$    0.82
 & 
    0.88
$\pm$    0.02
 & 
   32.83
$\pm$    0.94
 & 
   59.31
 & 
    3.49
$\pm$    0.02
 & 
  121.23
 & 
   82.00
 & 
   31.66
 \\
TEX 1156+213   
	& 
   14.69
$\pm$    0.70
 & 
    1.11
$\pm$    0.02
 & 
   45.00
$\pm$    0.94
 & 
  116.09
 & 
    7.91
$\pm$    0.02
 & 
  301.97
 & 
   58.40
 & 
   17.46
 \\
\hline
\end{tabular}
}
\end{minipage}
\end{table*}

\begin{table*}
\begin{minipage}{14cm}
\renewcommand{\thefootnote}{\alph{footnote}}
\caption{Physical measurements}
\label{tab:phys}
\scalebox{0.75}{
\begin{tabular}{lccccccc}
Object  & log $\lambda\,L_{\lambda}(1450 \textrm{\AA)}$  & log $\lambda\,L_{\lambda}(5100 \textrm{\AA)}$ & $\sigma_{l,\textrm{\CIV}}$ & $\sigma_{l,\textrm{H}\beta}$ & \fwhmciv    & \fwhmhb     & FWHM$_{\lambda1400}$ \\
        & ergs s$^{-1}$                                  & ergs s$^{-1}$                                 & km s$^{-1}$           & km s$^{-1}$         & km s$^{-1}$ & km s$^{-1}$ & km s$^{-1}$\\
\hline
3C 110   
	& 
   46.43
 & 
   45.99
 & 
        5260
$\pm$         789
 & 
        5431
$\pm$         814
 & 
        5700
$\pm$         855
 & 
       12450
$\pm$        1867
 & 
       13817
$\pm$        2072
 \\
3C 175   
	& 
   46.34
 & 
   46.11
 & 
        5322
$\pm$         798
 & 
        9599
$\pm$        1439
 & 
        6915
$\pm$        1037
 & 
       20925
$\pm$        3138
 & 
       17544
$\pm$        2631
 \\
3C 186   
	& 
   46.07
 & 
   45.79
 & 
        4521
$\pm$         678
 & 
\nodata

 & 
        6290
$\pm$         943
 & 
\nodata

 & 
       14101
$\pm$        2115
 \\
3C 207   
	& 
   45.65
 & 
   45.48
 & 
        3386
$\pm$         507
 & 
        5342
$\pm$         801
 & 
        4935
$\pm$         740
 & 
        3505
$\pm$         525
 & 
        4992
$\pm$         748
 \\
3C 215   
	& 
   45.02
 & 
   44.94
 & 
        5049
$\pm$         757
 & 
        3475
$\pm$         521
 & 
        5605
$\pm$         840
 & 
        6760
$\pm$        1014
 & 
        4459
$\pm$         668
 \\
3C 232   
	& 
   45.87
 & 
   45.72
 & 
        4053
$\pm$         608
 & 
        3328
$\pm$         499
 & 
        7145
$\pm$        1071
 & 
        4655
$\pm$         698
 & 
        6153
$\pm$         923
 \\
3C 254   
	& 
   45.63
 & 
   45.36
 & 
        5105
$\pm$         765
 & 
        7142
$\pm$        1071
 & 
        5205
$\pm$         780
 & 
       14095
$\pm$        2114
 & 
        6305
$\pm$         945
 \\
3C 263   
	& 
   46.33
 & 
   46.01
 & 
        4483
$\pm$         672
 & 
        4041
$\pm$         606
 & 
        3310
$\pm$         496
 & 
        4970
$\pm$         745
 & 
        5025
$\pm$         753
 \\
3C 277.1   
	& 
   45.04
 & 
   44.67
 & 
        3333
$\pm$         500
 & 
        3051
$\pm$         457
 & 
        3215
$\pm$         482
 & 
        3835
$\pm$         575
 & 
        4339
$\pm$         650
 \\
3C 281   
	& 
   45.68
 & 
   45.32
 & 
        5184
$\pm$         777
 & 
        4579
$\pm$         686
 & 
        4865
$\pm$         729
 & 
        7985
$\pm$        1197
 & 
       12003
$\pm$        1800
 \\
3C 288.1   
	& 
   46.01
 & 
   45.62
 & 
        4780
$\pm$         717
 & 
        3906
$\pm$         585
 & 
        4015
$\pm$         602
 & 
        8970
$\pm$        1345
 & 
       12217
$\pm$        1832
 \\
3C 334   
	& 
   46.00
 & 
   45.58
 & 
        3983
$\pm$         597
 & 
        4722
$\pm$         708
 & 
        5745
$\pm$         861
 & 
        6345
$\pm$         951
 & 
       12737
$\pm$        1910
 \\
3C 37   
	& 
   45.28
 & 
   44.89
 & 
        3225
$\pm$         483
 & 
        3893
$\pm$         584
 & 
        3360
$\pm$         504
 & 
        4280
$\pm$         642
 & 
        4619
$\pm$         692
 \\
3C 446   
	& 
   46.19
 & 
   46.43
 & 
        3780
$\pm$         567
 & 
\nodata

 & 
        3390
$\pm$         508
 & 
\nodata

 & 
       12181
$\pm$        1827
 \\
3C 47   
	& 
   45.27
 & 
   44.88
 & 
        5265
$\pm$         789
 & 
        6892
$\pm$        1033
 & 
        5450
$\pm$         817
 & 
       14005
$\pm$        2100
 & 
        4415
$\pm$         662
 \\
4C 01.04   
	& 
   44.43
 & 
   44.69
 & 
        4553
$\pm$         683
 & 
        4196
$\pm$         629
 & 
        6665
$\pm$         999
 & 
        9905
$\pm$        1485
 & 
       12310
$\pm$        1846
 \\
4C 06.69   
	& 
   46.73
 & 
   46.43
 & 
        4288
$\pm$         643
 & 
        2340
$\pm$         351
 & 
        5620
$\pm$         843
 & 
        4015
$\pm$         602
 & 
       10807
$\pm$        1621
 \\
4C 10.06   
	& 
   45.85
 & 
   45.47
 & 
        4566
$\pm$         684
 & 
        5922
$\pm$         888
 & 
        3785
$\pm$         567
 & 
        4735
$\pm$         710
 & 
        8366
$\pm$        1254
 \\
4C 11.69   
	& 
   46.51
 & 
   46.38
 & 
        3527
$\pm$         529
 & 
\nodata

 & 
        3185
$\pm$         477
 & 
\nodata

 & 
       12207
$\pm$        1831
 \\
4C 12.40   
	& 
   45.53
 & 
   45.13
 & 
        4485
$\pm$         672
 & 
        2329
$\pm$         349
 & 
        5300
$\pm$         795
 & 
        3565
$\pm$         534
 & 
       12378
$\pm$        1856
 \\
4C 19.44   
	& 
   46.30
 & 
   45.99
 & 
        3624
$\pm$         543
 & 
        3152
$\pm$         472
 & 
        2730
$\pm$         409
 & 
        4575
$\pm$         686
 & 
        4367
$\pm$         655
 \\
4C 20.24   
	& 
   46.22
 & 
   45.95
 & 
        3654
$\pm$         548
 & 
\nodata

 & 
        3525
$\pm$         528
 & 
\nodata

 & 
        4477
$\pm$         671
 \\
4C 22.26   
	& 
   45.83
 & 
   45.52
 & 
        4390
$\pm$         658
 & 
\nodata

 & 
        5015
$\pm$         752
 & 
\nodata

 & 
        5931
$\pm$         889
 \\
4C 30.25   
	& 
   45.64
 & 
   45.13
 & 
        3715
$\pm$         557
 & 
\nodata

 & 
        3730
$\pm$         559
 & 
\nodata

 & 
        4806
$\pm$         720
 \\
4C 31.63   
	& 
   46.21
 & 
   45.61
 & 
        4322
$\pm$         648
 & 
        2699
$\pm$         404
 & 
        4840
$\pm$         726
 & 
        3395
$\pm$         509
 & 
        5033
$\pm$         754
 \\
4C 34.47   
	& 
   45.27
 & 
   44.97
 & 
        3785
$\pm$         567
 & 
        3277
$\pm$         491
 & 
        2855
$\pm$         428
 & 
        5015
$\pm$         752
 & 
        4159
$\pm$         623
 \\
4C 39.25   
	& 
   46.25
 & 
   45.97
 & 
        4429
$\pm$         664
 & 
        3538
$\pm$         530
 & 
        4775
$\pm$         716
 & 
        6400
$\pm$         960
 & 
       11040
$\pm$        1656
 \\
4C 40.24   
	& 
   45.96
 & 
   45.74
 & 
        3265
$\pm$         489
 & 
\nodata

 & 
        4920
$\pm$         738
 & 
\nodata

 & 
        3973
$\pm$         595
 \\
4C 41.21   
	& 
   46.27
 & 
   45.64
 & 
        3736
$\pm$         560
 & 
        3170
$\pm$         475
 & 
        3800
$\pm$         570
 & 
        3445
$\pm$         516
 & 
        6075
$\pm$         911
 \\
4C 49.22   
	& 
   45.21
 & 
   44.84
 & 
        3465
$\pm$         519
 & 
        2741
$\pm$         411
 & 
        4535
$\pm$         680
 & 
        3910
$\pm$         586
 & 
        5151
$\pm$         772
 \\
4C 55.17   
	& 
   45.83
 & 
   45.72
 & 
        3777
$\pm$         566
 & 
\nodata

 & 
        6420
$\pm$         963
 & 
\nodata

 & 
        9769
$\pm$        1465
 \\
4C 58.29   
	& 
   46.73
 & 
   46.42
 & 
        4211
$\pm$         631
 & 
\nodata

 & 
        5745
$\pm$         861
 & 
\nodata

 & 
       12128
$\pm$        1819
 \\
4C 64.15   
	& 
   46.14
 & 
   45.98
 & 
        5230
$\pm$         784
 & 
\nodata

 & 
        7245
$\pm$        1086
 & 
\nodata

 & 
       12366
$\pm$        1854
 \\
4C 73.18   
	& 
   45.80
 & 
   45.55
 & 
        3585
$\pm$         537
 & 
        2704
$\pm$         405
 & 
        3560
$\pm$         534
 & 
        3095
$\pm$         464
 & 
        4172
$\pm$         625
 \\
B2 0742+31   
	& 
   45.89
 & 
   45.77
 & 
        4475
$\pm$         671
 & 
        5203
$\pm$         780
 & 
        4890
$\pm$         733
 & 
       10690
$\pm$        1603
 & 
        5623
$\pm$         843
 \\
B2 1351+31   
	& 
   46.07
 & 
   45.81
 & 
        4728
$\pm$         709
 & 
\nodata

 & 
        3690
$\pm$         553
 & 
\nodata

 & 
       12220
$\pm$        1833
 \\
B2 1555+33   
	& 
   45.57
 & 
   45.39
 & 
        3785
$\pm$         567
 & 
\nodata

 & 
        4240
$\pm$         636
 & 
\nodata

 & 
        6663
$\pm$         999
 \\
B2 1611+34   
	& 
   46.50
 & 
   46.29
 & 
        3836
$\pm$         575
 & 
        3074
$\pm$         461
 & 
        4625
$\pm$         693
 & 
        4795
$\pm$         719
 & 
        6557
$\pm$         983
 \\
IRAS F07546+3928   
	& 
   44.76
 & 
   44.86
 & 
        3486
$\pm$         522
 & 
        2999
$\pm$         449
 & 
        3035
$\pm$         455
 & 
        2785
$\pm$         417
 & 
        3624
$\pm$         543
 \\
MC2 0042+101   
	& 
   44.99
 & 
   44.94
 & 
        4616
$\pm$         692
 & 
        3517
$\pm$         527
 & 
        4195
$\pm$         629
 & 
        8270
$\pm$        1240
 & 
        7893
$\pm$        1184
 \\
MC2 1146+111   
	& 
   45.66
 & 
   45.52
 & 
        3926
$\pm$         588
 & 
        3438
$\pm$         515
 & 
        3715
$\pm$         557
 & 
        7835
$\pm$        1175
 & 
\nodata

 \\
MRK 506   
	& 
   43.88
 & 
   43.77
 & 
        3563
$\pm$         534
 & 
        3248
$\pm$         487
 & 
        5290
$\pm$         793
 & 
        4840
$\pm$         726
 & 
        5408
$\pm$         811
 \\
MRK 509   
	& 
   44.51
 & 
   44.19
 & 
        3785
$\pm$         567
 & 
        2645
$\pm$         396
 & 
        4710
$\pm$         706
 & 
        3825
$\pm$         573
 & 
        4886
$\pm$         732
 \\
OS 562   
	& 
   46.21
 & 
   45.82
 & 
        4234
$\pm$         635
 & 
        2715
$\pm$         407
 & 
        3470
$\pm$         520
 & 
        3305
$\pm$         495
 & 
        7870
$\pm$        1180
 \\
PG 0052+251   
	& 
   45.27
 & 
   44.88
 & 
        4589
$\pm$         688
 & 
        3864
$\pm$         579
 & 
        5815
$\pm$         872
 & 
        6460
$\pm$         969
 & 
        5610
$\pm$         841
 \\
PG 0844+349   
	& 
   44.66
 & 
   44.39
 & 
        2880
$\pm$         432
 & 
        2010
$\pm$         301
 & 
        4550
$\pm$         682
 & 
        2870
$\pm$         430
 & 
        4607
$\pm$         691
 \\
PG 0947+396   
	& 
   45.15
 & 
   44.79
 & 
        3812
$\pm$         571
 & 
        2822
$\pm$         423
 & 
        3925
$\pm$         588
 & 
        4340
$\pm$         651
 & 
        4977
$\pm$         746
 \\
PG 0953+414   
	& 
   45.87
 & 
   45.38
 & 
        3239
$\pm$         485
 & 
        2429
$\pm$         364
 & 
        3810
$\pm$         571
 & 
        2990
$\pm$         448
 & 
        4213
$\pm$         631
 \\
PG 1001+054   
	& 
   44.66
 & 
   44.40
 & 
        3832
$\pm$         574
 & 
        1529
$\pm$         229
 & 
        3130
$\pm$         469
 & 
        2615
$\pm$         392
 & 
        3737
$\pm$         560
 \\
PG 1100+772   
	& 
   45.90
 & 
   45.44
 & 
        4784
$\pm$         717
 & 
        4158
$\pm$         623
 & 
        4775
$\pm$         716
 & 
        9390
$\pm$        1408
 & 
        3556
$\pm$         533
 \\
PG 1103-006   
	& 
   45.67
 & 
   45.30
 & 
        4212
$\pm$         631
 & 
        3187
$\pm$         478
 & 
        4515
$\pm$         677
 & 
        5270
$\pm$         790
 & 
        4964
$\pm$         744
 \\
PG 1114+445   
	& 
   44.75
 & 
   44.80
 & 
        3942
$\pm$         591
 & 
        3789
$\pm$         568
 & 
        3935
$\pm$         590
 & 
        4825
$\pm$         723
 & 
        6283
$\pm$         942
 \\
PG 1115+407   
	& 
   45.11
 & 
   44.61
 & 
        4049
$\pm$         607
 & 
        1838
$\pm$         275
 & 
        4585
$\pm$         687
 & 
        1895
$\pm$         284
 & 
        5073
$\pm$         761
 \\
PG 1116+215   
	& 
   45.69
 & 
   45.10
 & 
        3650
$\pm$         547
 & 
        2167
$\pm$         325
 & 
        3865
$\pm$         579
 & 
        2975
$\pm$         446
 & 
        5682
$\pm$         852
 \\
PG 1202+281   
	& 
   44.48
 & 
   44.50
 & 
        3932
$\pm$         589
 & 
        5100
$\pm$         765
 & 
        2945
$\pm$         441
 & 
        4950
$\pm$         742
 & 
        4882
$\pm$         732
 \\
PG 1216+069   
	& 
   45.62
 & 
   45.48
 & 
        5141
$\pm$         771
 & 
        3376
$\pm$         506
 & 
        3105
$\pm$         465
 & 
        5950
$\pm$         892
 & 
        6216
$\pm$         932
 \\
PG 1226+023   
	& 
   46.45
 & 
   45.97
 & 
        3724
$\pm$         558
 & 
        2791
$\pm$         418
 & 
        4530
$\pm$         679
 & 
        3405
$\pm$         510
 & 
        4571
$\pm$         685
 \\
PG 1259+593   
	& 
   46.26
 & 
   45.69
 & 
        4105
$\pm$         615
 & 
        3664
$\pm$         549
 & 
        6880
$\pm$        1032
 & 
        4035
$\pm$         605
 & 
        5311
$\pm$         796
 \\
PG 1309+355   
	& 
   45.21
 & 
   45.07
 & 
        4775
$\pm$         716
 & 
        4632
$\pm$         694
 & 
        2815
$\pm$         422
 & 
        3640
$\pm$         546
 & 
        4246
$\pm$         636
 \\
PG 1322+659   
	& 
   45.07
 & 
   44.70
 & 
        3510
$\pm$         526
 & 
        1776
$\pm$         266
 & 
        3690
$\pm$         553
 & 
        3285
$\pm$         492
 & 
        4544
$\pm$         681
 \\
PG 1351+640   
	& 
   44.71
 & 
   44.90
 & 
        2953
$\pm$         442
 & 
        3355
$\pm$         503
 & 
        4050
$\pm$         607
 & 
        6205
$\pm$         930
 & 
        3790
$\pm$         568
 \\
PG 1352+183   
	& 
   44.96
 & 
   44.52
 & 
        3827
$\pm$         574
 & 
        3162
$\pm$         474
 & 
        3755
$\pm$         563
 & 
        4210
$\pm$         631
 & 
        5981
$\pm$         897
 \\
PG 1402+261   
	& 
   45.41
 & 
   44.84
 & 
        3894
$\pm$         584
 & 
        1992
$\pm$         298
 & 
        4550
$\pm$         682
 & 
        2100
$\pm$         315
 & 
        6397
$\pm$         959
 \\
PG 1411+442   
	& 
   44.86
 & 
   44.67
 & 
        2003
$\pm$         300
 & 
        3465
$\pm$         519
 & 
        2040
$\pm$         306
 & 
        2800
$\pm$         420
 & 
        3921
$\pm$         588
 \\
PG 1415+451   
	& 
   44.70
 & 
   44.45
 & 
        2656
$\pm$         398
 & 
        1950
$\pm$         292
 & 
        3725
$\pm$         558
 & 
        2560
$\pm$         384
 & 
        4711
$\pm$         706
 \\
PG 1425+267   
	& 
   45.54
 & 
   45.28
 & 
        5076
$\pm$         761
 & 
        6235
$\pm$         935
 & 
        7060
$\pm$        1059
 & 
        9875
$\pm$        1481
 & 
        6565
$\pm$         984
 \\
PG 1427+480   
	& 
   45.21
 & 
   44.74
 & 
        2758
$\pm$         413
 & 
        3975
$\pm$         596
 & 
        2835
$\pm$         425
 & 
        2405
$\pm$         360
 & 
        5796
$\pm$         869
 \\
PG 1440+356   
	& 
   44.97
 & 
   44.60
 & 
        1959
$\pm$         293
 & 
        1109
$\pm$         166
 & 
        2130
$\pm$         319
 & 
        1745
$\pm$         261
 & 
        3917
$\pm$         587
 \\
PG 1444+407   
	& 
   45.55
 & 
   45.17
 & 
        3428
$\pm$         514
 & 
        2037
$\pm$         305
 & 
        4425
$\pm$         663
 & 
        2750
$\pm$         412
 & 
        6224
$\pm$         933
 \\
PG 1512+370   
	& 
   45.55
 & 
   45.15
 & 
        5238
$\pm$         785
 & 
        5458
$\pm$         818
 & 
        3970
$\pm$         595
 & 
        7690
$\pm$        1153
 & 
        5628
$\pm$         844
 \\
PG 1534+580   
	& 
   43.61
 & 
   43.37
 & 
        3239
$\pm$         485
 & 
        2888
$\pm$         433
 & 
        3790
$\pm$         568
 & 
        4505
$\pm$         675
 & 
        4609
$\pm$         691
 \\
PG 1543+489   
	& 
   45.80
 & 
   45.38
 & 
        3842
$\pm$         576
 & 
        1491
$\pm$         223
 & 
        5625
$\pm$         843
 & 
        2320
$\pm$         348
 & 
        5424
$\pm$         813
 \\
PG 1545+210   
	& 
   45.39
 & 
   45.07
 & 
        4654
$\pm$         698
 & 
        3382
$\pm$         507
 & 
        4560
$\pm$         684
 & 
        6885
$\pm$        1032
 & 
        6148
$\pm$         922
 \\
PG 1626+554   
	& 
   45.06
 & 
   44.61
 & 
        4291
$\pm$         643
 & 
        2998
$\pm$         449
 & 
        3815
$\pm$         572
 & 
        4390
$\pm$         658
 & 
        6502
$\pm$         975
 \\
PG 1704+608   
	& 
   45.91
 & 
   45.72
 & 
        5340
$\pm$         801
 & 
        4637
$\pm$         695
 & 
        4015
$\pm$         602
 & 
       10465
$\pm$        1569
 & 
       11557
$\pm$        1733
 \\
PG 2214+139   
	& 
   44.84
 & 
   44.54
 & 
        3678
$\pm$         551
 & 
        3137
$\pm$         470
 & 
        2690
$\pm$         403
 & 
        5845
$\pm$         876
 & 
        6551
$\pm$         982
 \\
PG 2251+113   
	& 
   45.73
 & 
   45.53
 & 
        4234
$\pm$         635
 & 
        2996
$\pm$         449
 & 
        4805
$\pm$         720
 & 
        4060
$\pm$         609
 & 
        4660
$\pm$         699
 \\
PG 2349-014   
	& 
   45.09
 & 
   44.78
 & 
        4211
$\pm$         631
 & 
        3728
$\pm$         559
 & 
        5675
$\pm$         851
 & 
        6325
$\pm$         948
 & 
        4476
$\pm$         671
 \\
PKS 0112-017   
	& 
   46.44
 & 
   46.07
 & 
        4019
$\pm$         602
 & 
\nodata

 & 
        5030
$\pm$         754
 & 
\nodata

 & 
        8572
$\pm$        1285
 \\
PKS 0403-13   
	& 
   45.69
 & 
   45.53
 & 
        3977
$\pm$         596
 & 
        2856
$\pm$         428
 & 
        3325
$\pm$         498
 & 
        3735
$\pm$         560
 & 
        4577
$\pm$         686
 \\
PKS 0859-14   
	& 
   46.68
 & 
   46.45
 & 
        3737
$\pm$         560
 & 
        5260
$\pm$         789
 & 
        4520
$\pm$         678
 & 
        4615
$\pm$         692
 & 
        4737
$\pm$         710
 \\
PKS 1127-14   
	& 
   46.50
 & 
   46.21
 & 
        3501
$\pm$         525
 & 
\nodata

 & 
        3695
$\pm$         554
 & 
\nodata

 & 
       12263
$\pm$        1839
 \\
PKS 1656+053   
	& 
   46.44
 & 
   46.26
 & 
\nodata

 & 
        2832
$\pm$         424
 & 
\nodata

 & 
        3510
$\pm$         526
 & 
        6933
$\pm$        1040
 \\
PKS 2216-03   
	& 
   46.38
 & 
   46.31
 & 
        4504
$\pm$         675
 & 
        2962
$\pm$         444
 & 
        3600
$\pm$         540
 & 
        4415
$\pm$         662
 & 
        5580
$\pm$         837
 \\
TEX 1156+213   
	& 
   45.36
 & 
   45.00
 & 
        5129
$\pm$         769
 & 
        3604
$\pm$         540
 & 
        3880
$\pm$         582
 & 
        7740
$\pm$        1161
 & 
        6156
$\pm$         923
 \\
\hline
\end{tabular}
}
\end{minipage}
\end{table*}

%ANALYSIS
%%%%%%%%%%%%%%%%%%%%%%%%%%%%%%%%%%%%%%%%%%%%%%%%%%%%%%%%%%%%%%%%%%%%%%%%%%%%%%%%%
\section{Analysis}
\label{sec:analysis}
%the effect for line width
\subsection{The Eigenvector 1 bias in \CIV}
%The most reliable black hole estimates come from the \Hb\ line, where the line width is determined predominately by the black hole mass.  However, in \CIV\ the line width depends at the very least on the black hole mass and some other as yet unknown driver indicated by EV1.  The FHWM measure of line width is particularly sensitive to the contamination in the \CIV\ line, introducing scatter between \CIV\ and \Hb\ black hole masses.  By estimating the strength of the emission from low-velocity gas that does not respond to the gravity of the black hole (e.g., the narrow ILR core component of the line), we can remove the contamination to the \CIV\ line and calculate a black hole mass that better agrees with \Hb-based estimates.
The trend of the FWHM of \CIV\ with the ratio Peak($\lambda$1400/\CIV) illustrated by \citet{wills93b} suggests that both EV1 and black hole mass play a role in determining the \CIV\ line profile.  Because this effect is not present in \Hb, we expect that it contributes significantly to the disagreement between velocity widths measured from \Hb\ and \CIV.

\begin{figure}
\begin{center}
\includegraphics[width=8.9 truecm]{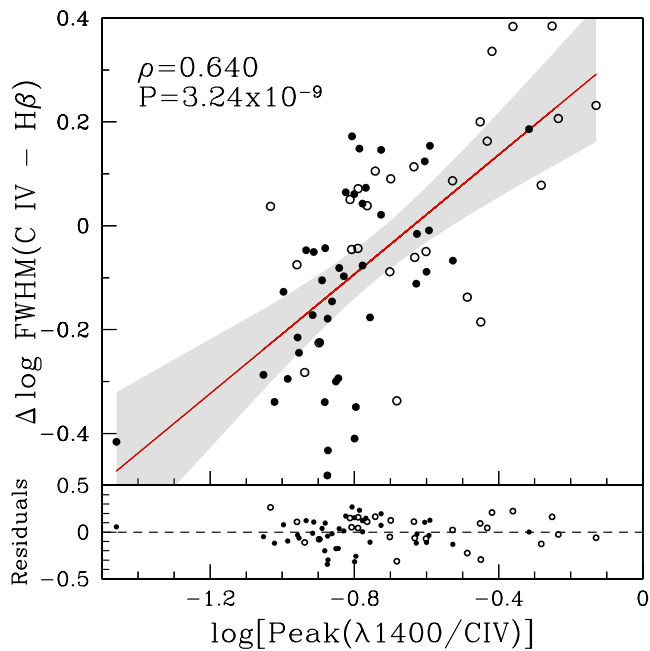}
\end{center}
\caption{Residuals in FWHM for \CIV\ and \Hb\ versus Peak($\lambda$1400/\CIV) (top) and residuals in the fit (bottom).  The Spearman Rank correlation coefficient and associated probability that this distribution of points occurs by chance are listed in the upper left corner.  The significant correlation indicates that the disagreement between \CIV\ and \Hb\ line widths that we hope to reduce depends on Peak($\lambda$1400/\CIV).  The fitted line is given in red with the shaded region indicating the 95\% confidence interval.  Radio-loud objects are indicated by solid circles and open circles are radio-quiet.}
\label{fig:effect}
\end{figure}

Figure~\ref{fig:effect} demonstrates that the ratio Peak($\lambda$1400/\CIV) is correlated with the significant scatter between FWHM of \Hb\ and \CIV.  We distinguish between RL and RQ points in the figure and find that they do not clearly populate separate relationships, though they do tend to occupy opposite ends of EV1.  

We performed a regression analysis to describe the dependence of the FWHM residuals on Peak($\lambda$1400/\CIV) and determined an expression for predicting the \Hb\ FWHM.  Line fitting in this investigation is done with an ordinary least-squares Y on X (OLS(Y$|$X)) fit.  The OLS(Y$|$X) method is to be used in instances where X is used to predict Y \citep{isobe90}, as is the case for our analysis.  In the OLS(Y$|$X) fit, uncertainties are used only in the Y direction so uncertainties on the peak ratio do not enter into the fit.  The assumption that the errors in peak ratio are small certainly seems reasonable given the values in the peak fluxes in Table~\ref{tab:EV1}, however if the systematic errors dominate over the errors in the fitting procedure these may be underestimated.  Because the velocity width measurements have equal percent uncertainties, the points are all weighted equally in the fit.  

In order to determine the uncertainty in the slope and intercept used to predict the FWHM of \Hb, we employ the model-independent Monte Carlo simulation used in \citet{peterson98}.  This method simulates uncertainties associated with the measurement uncertainties and with the individual points in the sample.  For our observed dataset of $N=69$ objects we do the following.  First, generate $N$ Y values within the 1$\sigma$ Gaussian uncertainties around the measured values to form a dataset.  Then, similar to the ``bootstrapping'' technique, $N$ sources are selected randomly from the synthetic dataset, without regard for whether a point has been previously selected.  Duplicate points that are sampled multiple times are then discarded from the sample, typically cutting the sample down by the Poisson probability of not selecting any given point, or $\sim1/e\approx37\%$.  After duplicates are discarded, the dataset is fit following the procedure used on the actual data.  This process is realized $10^{5}$ times to build a distribution in the slope and intercept parameters.  The standard deviation of these distributions are taken to be the uncertainties on the slope and intercept.

Equation~\ref{eqn:EV1} is the resulting method for estimating the FWHM of the \Hb\ line based on UV spectral measurements:

%the corrections
{\scriptsize
\begin{eqnarray}
\label{eqn:EV1}
\nonumber {\rm log}\left[\frac{\textrm{\fwhmhbpred}}{\textrm{km s}^{-1}}\right] &=& \rm{log}\left[\frac{\textrm{\fwhmciv}}{\textrm{km s}^{-1}}\right] - (   0.366\pm   0.048) \\
&-& (   0.574\pm   0.061)\,\rm{log}\left[\textrm{Peak}\left(\frac{\lambda1400}{\textrm{\CIV}}\right)\right].
\end{eqnarray}}

This expression treats the RL and RQ sources together.  We have investigated the relationships for RL and RQ objects separately and find that they are not significantly different within the 95\% confidence intervals, as is suggested by the distribution of RL and RQ points in Figure~\ref{fig:effect}.  We note that, in the case of the RL sources, the correlation appears to be driven by two outliers in the peak ratio.  A Monte Carlo bootstrapping simulation that would likely reveal any sample dependence of RL correlation is beyond the scope of this paper, but we do wish to draw the readers attention to this possible caveat.

\begin{figure*}
\begin{minipage}[!b]{8cm}
\centering
\includegraphics[width=8cm]{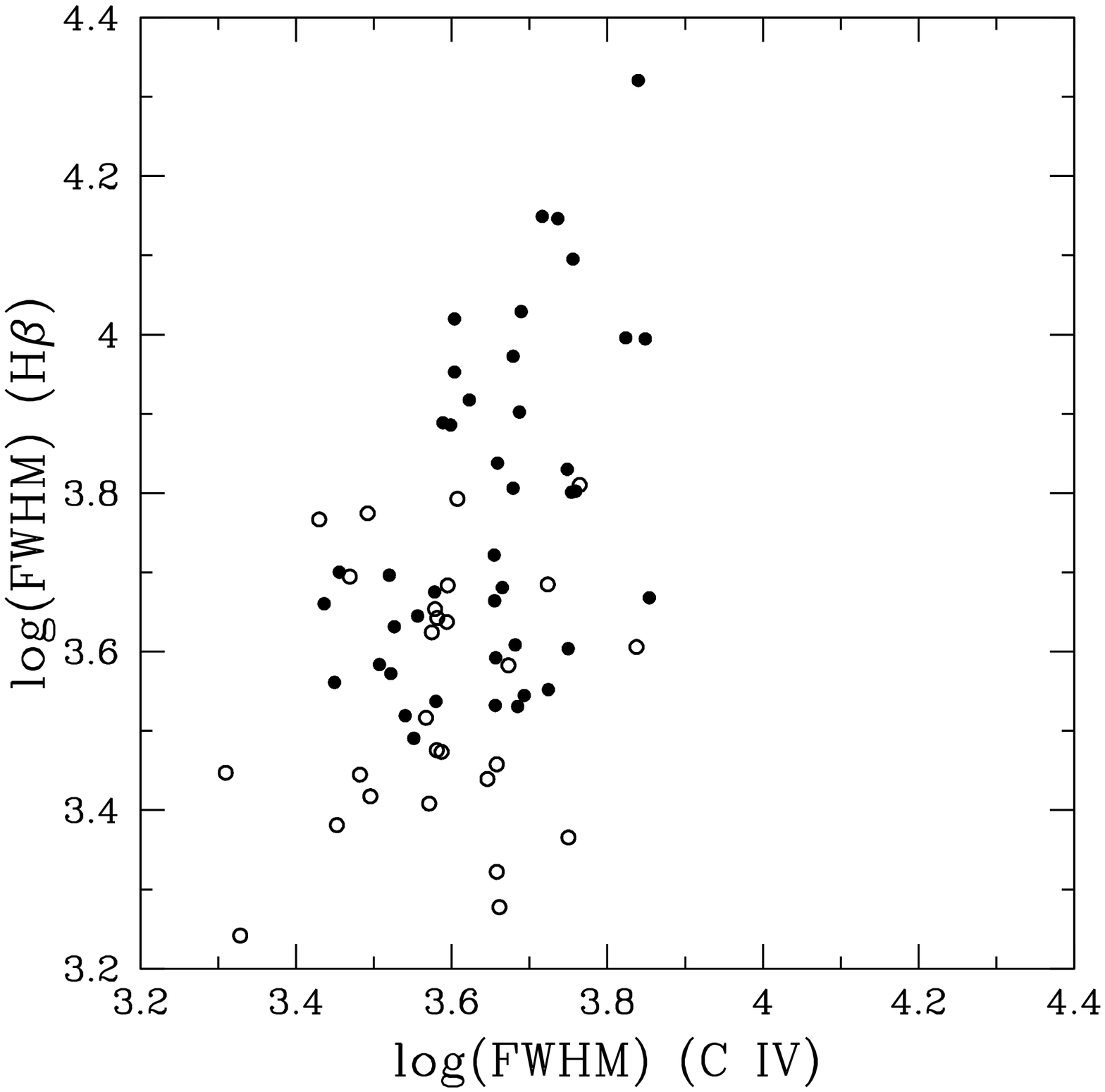}
\end{minipage}\hspace{0.6cm}
\hspace{0.6cm}
\begin{minipage}[!b]{8cm}
\centering
\includegraphics[width=8cm]{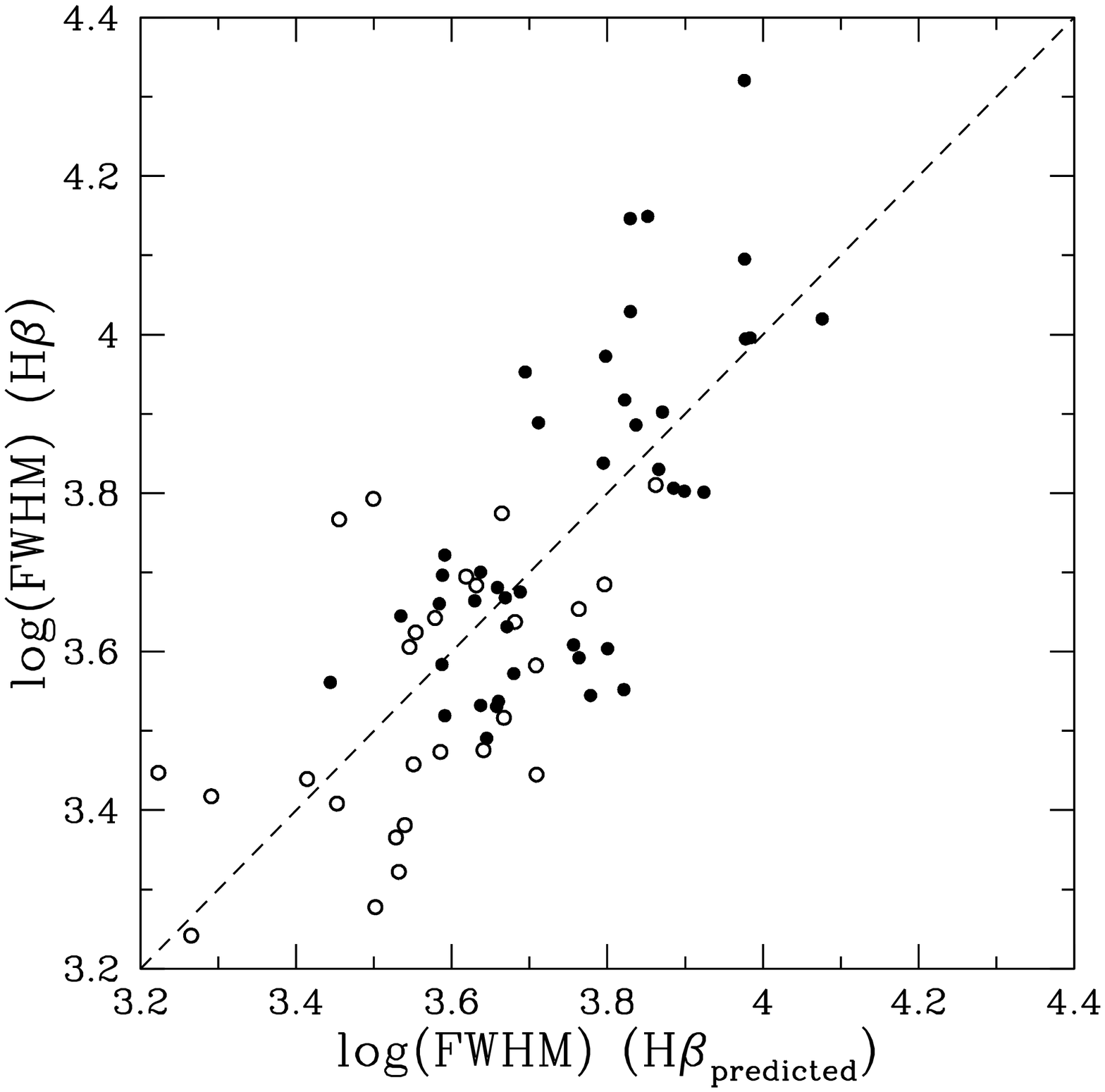}
\end{minipage}
\hspace{0.6cm}                   
\caption{The FWHM of \Hb\ versus \CIV\ (left panel) and versus the FWHM predicted for \Hb\ based on UV spectral information (right panel).  In the left panel, we do not expect the line widths to be equal and rather illustrate the scatter between \Hb\ and \CIV\ measurements.  In the right panel, we have used Equation~\ref{eqn:EV1} to predict the \Hb\ line width and expect the data to fall around the dashed one-to-one line.  The scatter decreases from 0.20 to 0.15 dex after including a Peak($\lambda$1400/\CIV) term and the points are fairly evenly distributed around the one-to-one line.  Radio-loud objects are indicated by solid circles and open circles are radio-quiet.}
\label{fig:FWHMcorr}
\end{figure*}

The FWHM predicted for \Hb\ based on \CIV\ and $\lambda1400$ shows less scatter with the true FWHM of \Hb\ than does \CIV\ alone.  Figure~\ref{fig:FWHMcorr} shows the relationship between \Hb\ and \CIV\ FWHM compared to \Hb\ and the predicted \Hb\ from Equation~\ref{eqn:EV1}.  Independent of the assumed relationship between FWHM of \CIV\ and \Hb, the initial scatter between them is 0.20 dex.  This is reduced to 0.15 dex and the points become fairly well centered around the one-to-one relationship by using Equation~\ref{eqn:EV1}.  This scatter reduction is in the dispersion of the distribution of the log(FWHM) residuals only, and is independent of the assuming a relationship between \CIV\ and \Hb\ FWHM.  Our goal here is to be able to effectively predict the \Hb\ FWHM, thus the reduction in scatter allows a more precise measurement.

%propagating that to Mbh
\subsection{Improved black hole masses}
\begin{figure}
\begin{center}
\includegraphics[width=8.9 truecm]{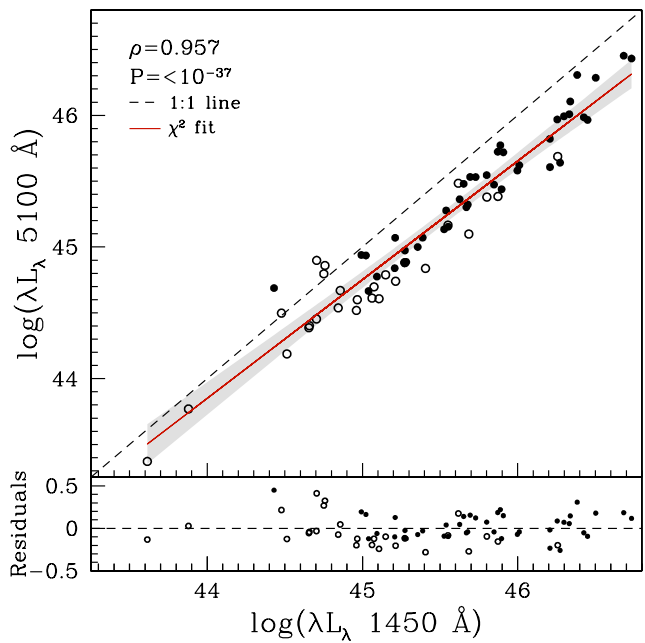}
\end{center}
\caption{5100 \AA\ luminosity versus 1450 \AA\ luminosity (top) and residuals in the fit (bottom). The dashed black line indicates where the luminosities are the same and the solid red line is the fitted line with the shaded region indicating the 95\% confidence interval.  Spearman Rank correlation coefficient and associated probability of finding this distribution of points by chance are listed in the upper left corner.  Radio-loud objects are indicated by solid circles and open circles are radio-quiet.}
\label{fig:L}
\end{figure}

%luminosity
In order to propagate the predicted \Hb\ FWHM into a black hole mass estimate, we derived a relationship for predicting the 5100 \AA\ luminosity from the 1450 \AA\ luminosity.  Figure~\ref{fig:L} shows that $\lambda\, L_{\lambda}$(5100 \AA) and $\lambda\, L_{\lambda}$(1450 \AA) are very significantly correlated, indicating that the 1450 \AA\ luminosity is a good proxy for the 5100 \AA\ luminosity.  When fitting the luminosities, we weighted each point equally.  The fitted line, determined using the same OLS(Y$|$X) fitting procedure, shown in Figure~\ref{fig:L} is:

{\footnotesize
\begin{eqnarray}
\label{eqn:L}
\nonumber {\rm log}\left[\frac{\lambda\textrm{L}_{\lambda}(5100 \rm{\AA})_{\rm{predicted}}}{\textrm{ergs s}^{-1}}\right] &=& (   0.901\pm   0.028)\,{\rm log}\left[\frac{\lambda\textrm{L}_{\lambda}(1450 \rm{\AA})}{\textrm{ergs s}^{-1}}\right] \\
&+& (   4.198\pm   1.265).
\end{eqnarray}}

The uncertainties on the slope and intercept are calculated via the same Monte Carlo method, assuming 5\% error in the luminosities consistent with \citet{shang07}.

\begin{figure*}
\begin{minipage}[!b]{8cm}
\centering
\includegraphics[width=8cm]{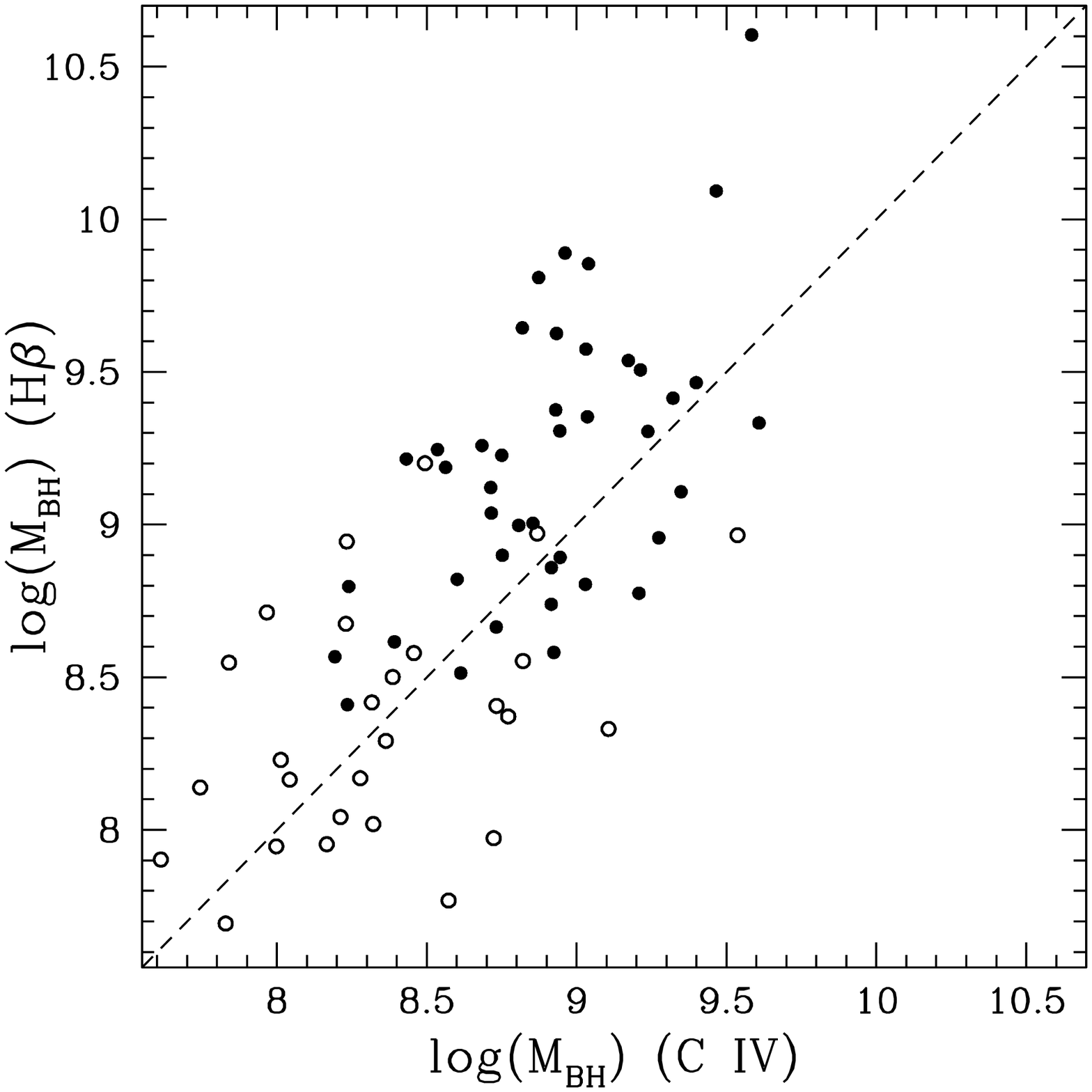}
\end{minipage}\hspace{0.6cm}
\hspace{0.6cm}
\begin{minipage}[!b]{8cm}
\centering
\includegraphics[width=8cm]{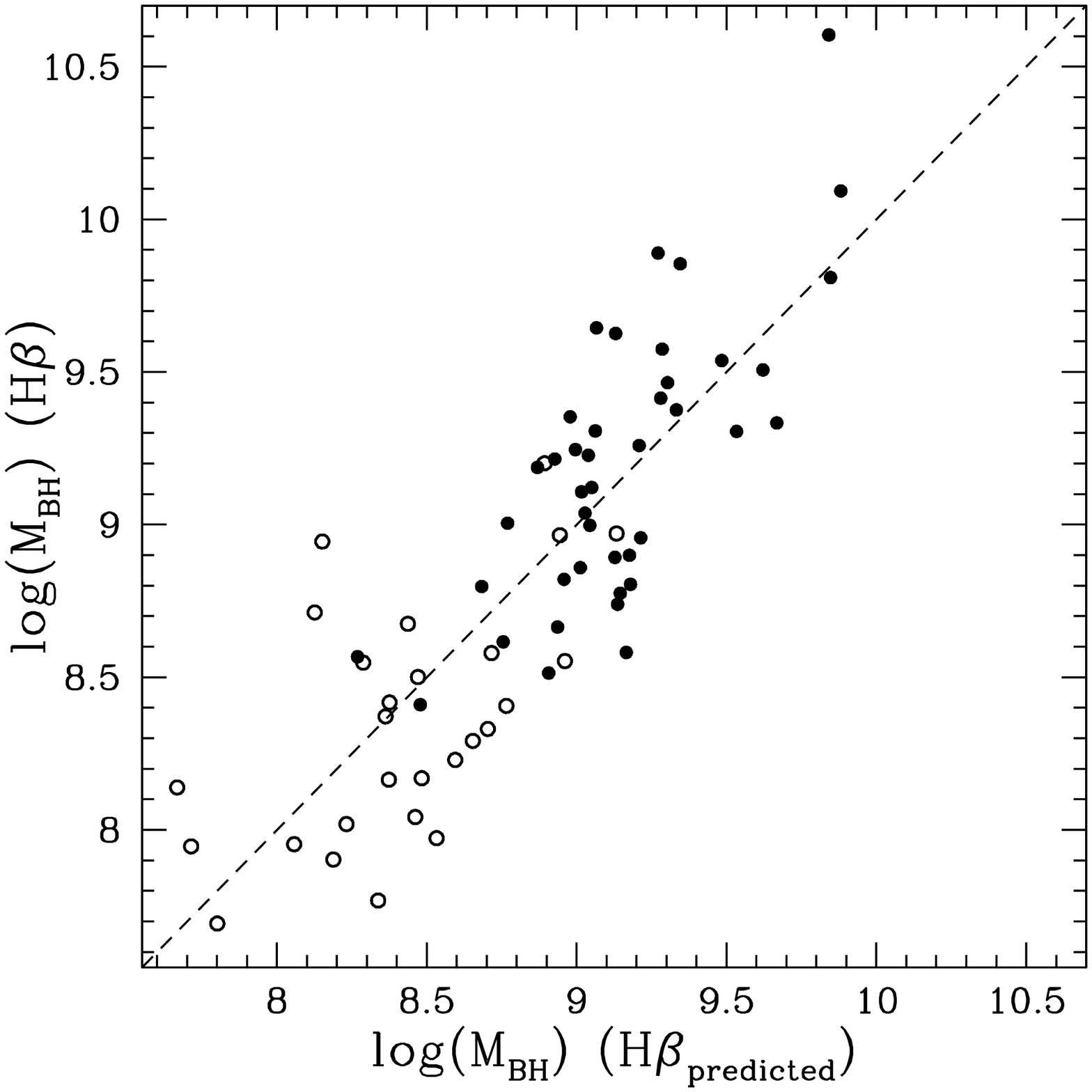}
\end{minipage}
\hspace{0.6cm}                   
\caption{\Hb-based black hole masses versus \CIV-based black hole masses (left panel) and the mass calculated from the predicted \Hb\ based on UV spectral information (right panel).  The one-to-one line is shown.  The scatter, originally 0.43 dex in the left panel, is reduced to 0.33 dex by including Peak($\lambda$1400/\CIV) and Equations~\ref{eqn:EV1} and \ref{eqn:L} when calculating black hole mass from spectral information in the \CIV\ region.  Radio-loud objects are indicated by solid circles and open circles are radio-quiet.}
\label{fig:Mbhcorr}
\end{figure*}

With the ability to predict the 5100 \AA\ luminosity and \Hb\ line width from UV spectral parameters, comes the possibility of better predicting the black hole mass using spectral properties in the \CIV\ wavelength region.  We calculated rehabilitated \CIV\ masses from the predicted FWHM of \Hb\ using Equation~\ref{eqn:EV1}, the predicted 5100 \AA\ continuum luminosity using Equation~\ref{eqn:L}, and the \Hb\ scaling relationship from \citet{vestergaard06}.  We found better agreement between \Hb-based black hole masses and those calculated from \CIV\ and Peak($\lambda$1400/\CIV) than those calculated from \CIV\ alone.  Figure~\ref{fig:Mbhcorr} shows the before and after with \Hb\ versus \CIV\ and \Hb\ versus predicted \Hb-based black hole masses.  The improvement is significant, with a reduction in scatter from 0.43 dex to 0.33 dex.  This method is independent of the choice in scaling relationship; it is not necessary to use the \citet{vestergaard06} scaling relationship, any relation using FWHM \Hb\ and 5100 \AA\ continuum luminosity could be substituted.

\begin{figure}
\begin{center}
\includegraphics[width=8.9 truecm]{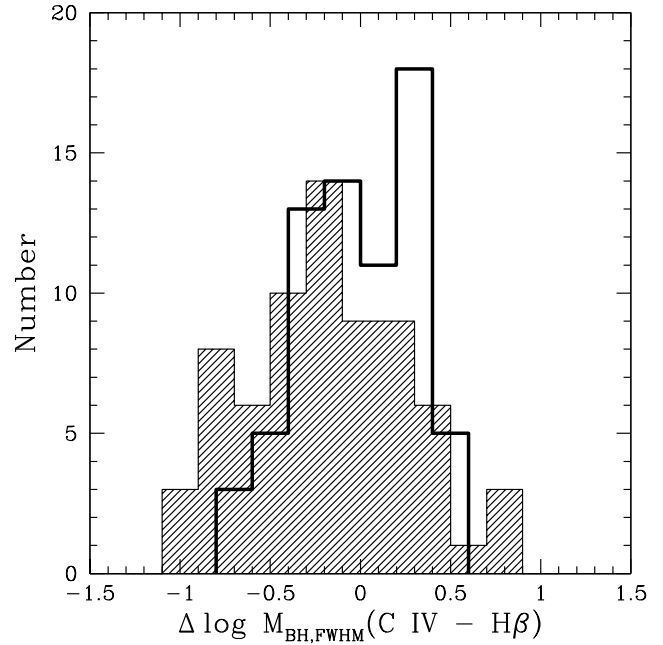}
\end{center}
\caption{Before and after histograms of the residuals between masses derived from FWHM.  The shaded histogram shows black hole mass residuals ``before'', using the difference between \CIV\ and \Hb-based masses.  The thick histogram shows black hole mass residuals ``after'', using the difference between predicted \Hb\ (using Equations~\ref{eqn:EV1} and \ref{eqn:L}) and \Hb-based masses.  The primary effect of predicting the FWHM of \Hb\ and using it to calculate black hole mass is a reduction in width of the distribution.}
\label{fig:fwhmhist}
\end{figure}

The reduction in scatter between \Hb\ and \CIV-based black hole masses can also be seen in the reduction in the width of the distribution of mass residuals.  Figure~\ref{fig:fwhmhist} shows ``before'' and ``after'' histograms for the FWHM-based black hole mass estimates.  It is clear that the scatter is reduced by the width reduction in the distribution.

\begin{figure}
\begin{center}
\includegraphics[width=8.9 truecm]{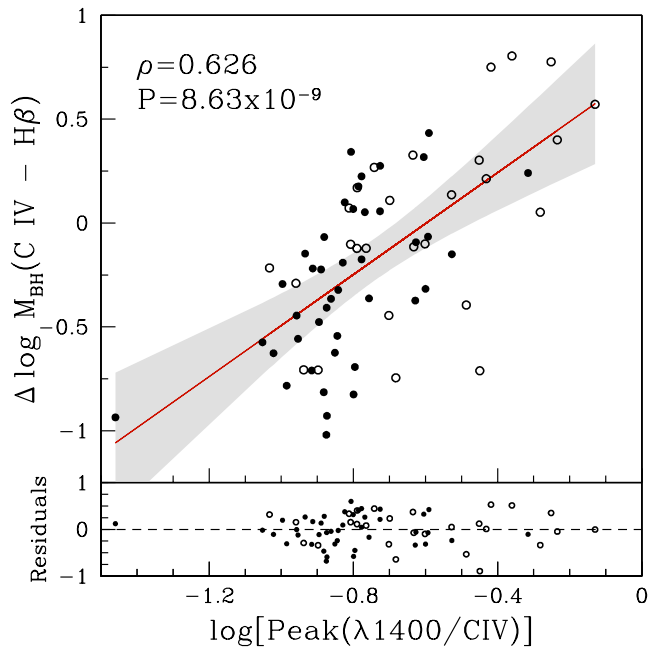}
\end{center}
\caption{Residuals in M$_{BH}$ for \CIV\ and \Hb\ versus the Peak($\lambda$1400/\CIV) (top) and residuals in the fit (bottom).  The Spearman Rank correlation coefficient and associated probability that this distribution of points occurs by chance are listed in the upper left corner.  The significant correlation indicates that there is a Peak($\lambda$1400/\CIV) dependence in the mass residuals similar to what was found for the FWHM.  The fitted line is given in red with the shaded region indicating the 95\% confidence interval.  Radio-loud objects are indicated by solid circles and open circles are radio-quiet.}
\label{fig:effectm}
\end{figure}

Black hole masses have already been calculated for thousands of objects using the \CIV\ prescription from \citet{vestergaard06}, for example in the catalog of \citet{shen11}.  In these cases, it is more straightforward to correct the pre-calculated black hole mass than to propagate through the predicted optical measurements.  We performed a regression analysis on the black hole mass residuals versus Peak($\lambda$1400/\CIV) (Figure~\ref{fig:effectm}).  We followed an identical fitting procedure here as for the FWHM residuals, with all of the uncertainty propagated from the velocity line widths and thus all points weighted equally in the fit.  Equation~\ref{eqn:M} gives the correction to pre-calculated black hole masses already derived from \CIV\ measurements:  

{\scriptsize
\begin{eqnarray}
\label{eqn:M}
\nonumber \textrm{log }\left[\frac{M_{BH}(\textrm{\Hb}_{\rm{predicted}})}{M_{\sun}}\right] &=& \textrm{log }\left[\frac{M_{BH}(\textrm{\CIV})}{M_{\sun}}\right] - (   0.734\pm   0.112) \\
&-& (   1.227\pm   0.136)\,\rm{log}\left[\textrm{Peak}\left(\frac{\lambda1400}{\textrm{\CIV}}\right)\right].
\end{eqnarray}}

This expression assumes the \CIV\ single-epoch scaling relationship of \citet{vestergaard06} and the resulting predicted \Hb\ black hole masses is akin to using the \Hb\ scaling relationship from the same authors.  The reduction in scatter achieved by using this correction is comparable to propagating the predicted optical measurements through the mass scaling relationship, although this method is less flexible because you cannot choose your mass scaling relationship.

A new mass scaling relationship for \CIV\ based on an updated reverberation mapping sample and methodology has recently become available from \citet{park13}.  We do not re-derive Equation~\ref{eqn:M} for the new scaling relationship since the motivation for a ready-made mass correction was backwards compatibility with catalogs like \citet{shen11}.  The effect of adopting this new mass scaling relationship in our sample is to increase the initial scatter between \Hb\ and \CIV-based masses from 0.43 to 0.45 dex.  Points in the left panel of Figure~\ref{fig:Mbhcorr} move to the left, generally farther from the one-to-one line and the scatter is increased.  Thus, the reduction in scatter achieved by correcting \citet{park13} \CIV-based masses with Equations~\ref{eqn:EV1} and \ref{eqn:L} is even more significant than for masses calculated from \citet{vestergaard06}.

%line dispersion
\subsection{An alternative measure of line width}
%more citations for using line dispersion?
%yue shen's article summarizes this
%peterson+(04)
Line dispersion is often used for calculating black hole masses instead of FWHM \citep{peterson04}.  Because of the contaminating emission in the \CIV\ line, line dispersion is often preferred for estimating black hole masses in high-redshift objects.  With detailed simulations, \citet{denney09a} show that line dispersion measurements are much less susceptible to contamination from low-velocity, non-virialized gas.  When a contaminating low-velocity component is strong in the \CIV\ line, the peak of the emission line shoots up although the base of the line remains largely unchanged.  A measurement of FWHM in such a line will give a value that is artificially narrow as the ``half maximum'' is much higher than it would otherwise would have been.  In these cases, the width of the contaminating component is being probed instead of the virialized base that is desired for black hole mass estimates.  

\begin{figure}
\begin{center}
\includegraphics[width=8.9 truecm]{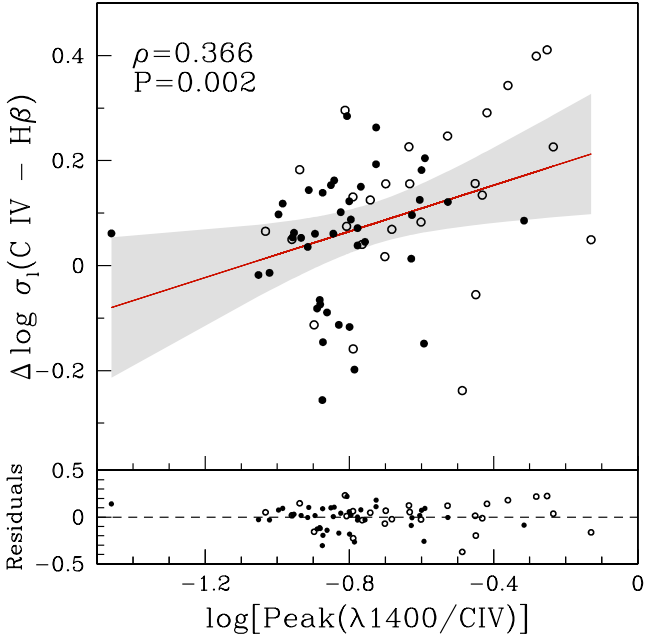}
\end{center}
\caption{Residuals in line dispersion for \CIV\ and \Hb\ versus Peak($\lambda$1400/\CIV).  The Spearman Rank correlation coefficient and associated probability that this distribution of points occurs by chance are listed in the upper left corner.  The correlation is significant at about the 3$\sigma$ level, much less than for FWHM.  Radio-loud objects are indicated by solid circles and open circles are radio-quiet.}
\label{fig:effects}
\end{figure}

Given this, it is not surprising that the residuals in line dispersion between the \CIV\ and \Hb\ emission lines are much less significantly correlated with Peak($\lambda$1400/\CIV) than for FWHM, although the correlation is still significant as seen in Figure~\ref{fig:effects}. 

The scatter in the relationship between line dispersion residuals and Peak($\lambda$1400/\CIV) make a correction less effective than for FWHM.  Equation~\ref{eqn:EV1sig} gives the predicted line dispersion for \Hb\ based on the \CIV\ line dispersion and Peak($\lambda$1400/\CIV) that was derived via the same OLS(Y$|$X) fitting procedure and Monte Carlo error simulation that are used throughout this work:

{\footnotesize
\begin{eqnarray}
\label{eqn:EV1sig}
\nonumber {\rm log}\left[\frac{\sigma_{l\textrm{,\CIV,predicted}}}{\textrm{km s}^{-1}}\right] &=& \rm{log}\left[\frac{\textrm{\siglciv}}{\textrm{km s}^{-1}}\right] - (   0.241\pm   0.055) \\
&-& (   0.220\pm   0.068)\,\rm{log}\left[\textrm{Peak}\left(\frac{\lambda1400}{\textrm{\CIV}}\right)\right].
\end{eqnarray}}

The reduction in scatter between \Hb\ and \CIV-based black hole masses estimated using line dispersion is minimal, on the order of 0.02 dex.  

\begin{figure}
\begin{center}
\includegraphics[width=8.9 truecm]{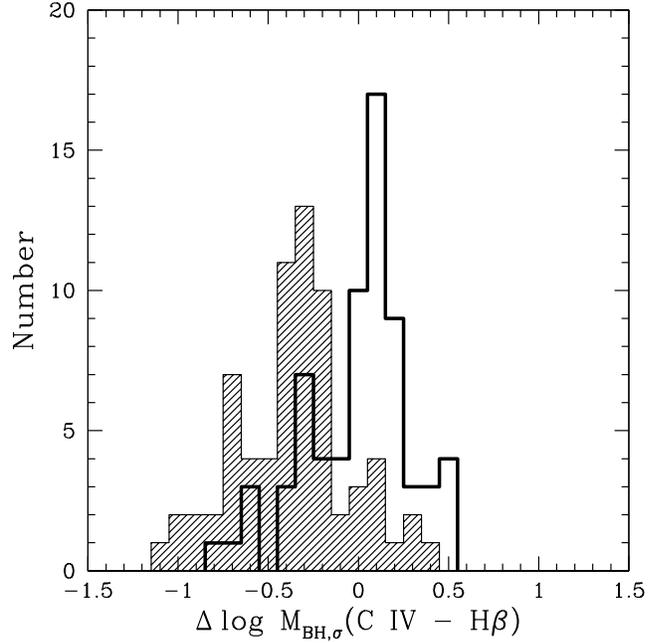}
\end{center}
\caption{Before and after histograms of the residuals between masses derived from line dispersion.  The shaded histogram shows black hole mass residuals ``before'', using the difference between \CIV\ and \Hb-based masses.  The thick histogram shows black hole mass residuals ``after'', using the difference between predicted \Hb\ (using Equations~\ref{eqn:L} and \ref{eqn:EV1sig}) and \Hb-based masses.  Predicting \Hb\ line dispersion shifts the distribution to zero, but does not decrease the width appreciably.}
\label{fig:sighist}
\end{figure}

Figure~\ref{fig:sighist} shows the distribution in mass residuals calculated from line dispersion for \CIV$-$\Hb\ and predicted \Hb$-$\Hb.  The distribution is initially relatively narrow, but offset from zero.  The result of predicting the \Hb\ mass from \CIV\ line dispersion is largely to remove that offset, the width of the distribution is not appreciably reduced.  We have attributed this offset to the Peak($\lambda$1400/\CIV) effect, but we note that there are other possibilities that we discuss further in Section~\ref{sec:discussion}.

%UV EV1 indicators
\subsection{Eigenvector 1 Indicators}
The correction we have derived is thus far is purely empirical, but the inclusion of Peak($\lambda$1400/\CIV) among the properties correlated in EV1 suggests an origin for the effect.  Eigenvector 1 was defined by \citet{bg92} based on measurements of optical spectral properties; variation in EV1 is often described by a continuum in the ratio EW(\FeII/\OIII).  Thus, in order to evaluate the source of the dependence of \Hb\ and \CIV\ line width residuals on Peak($\lambda$1400/\CIV), we investigate whether the effect holds for EW(\FeII/\OIII).  

\begin{figure}
\begin{center}
\includegraphics[width=8.9 truecm]{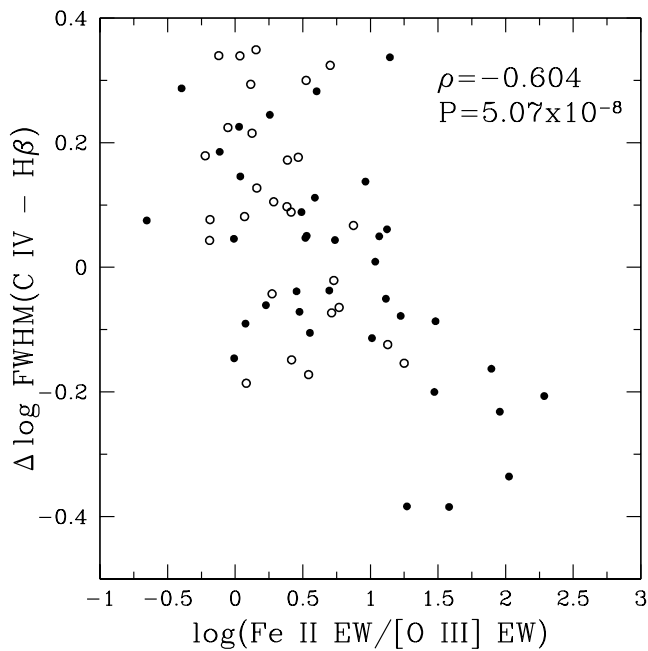}
\end{center}
\caption{Residuals in FWHM for \CIV\ and \Hb\ versus EW(\FeII/\OIII).  The Spearman Rank correlation coefficient and associated probability that this distribution of points occurs by chance are listed in the upper left corner.  The correlation is comparable in significance to that found for Peak($\lambda$1400/\CIV), indicating that the the correction derived here is an EV1 effect.  Radio-loud objects are indicated by solid circles and open circles are radio-quiet.}
\label{fig:optEV1effect}
\end{figure}

Figure~\ref{fig:optEV1effect} shows that the FWHM residuals depend strongly on EW(\FeII/\OIII), with a correlation that is comparable in significance to the one found for Peak($\lambda$1400/\CIV).  This indicates that it is indeed EV1 in addition to black hole mass that contributes to determining the velocity width of the \CIV\ line profile.

\begin{figure*}
\begin{minipage}[!b]{8cm}
\centering
\includegraphics[width=8cm]{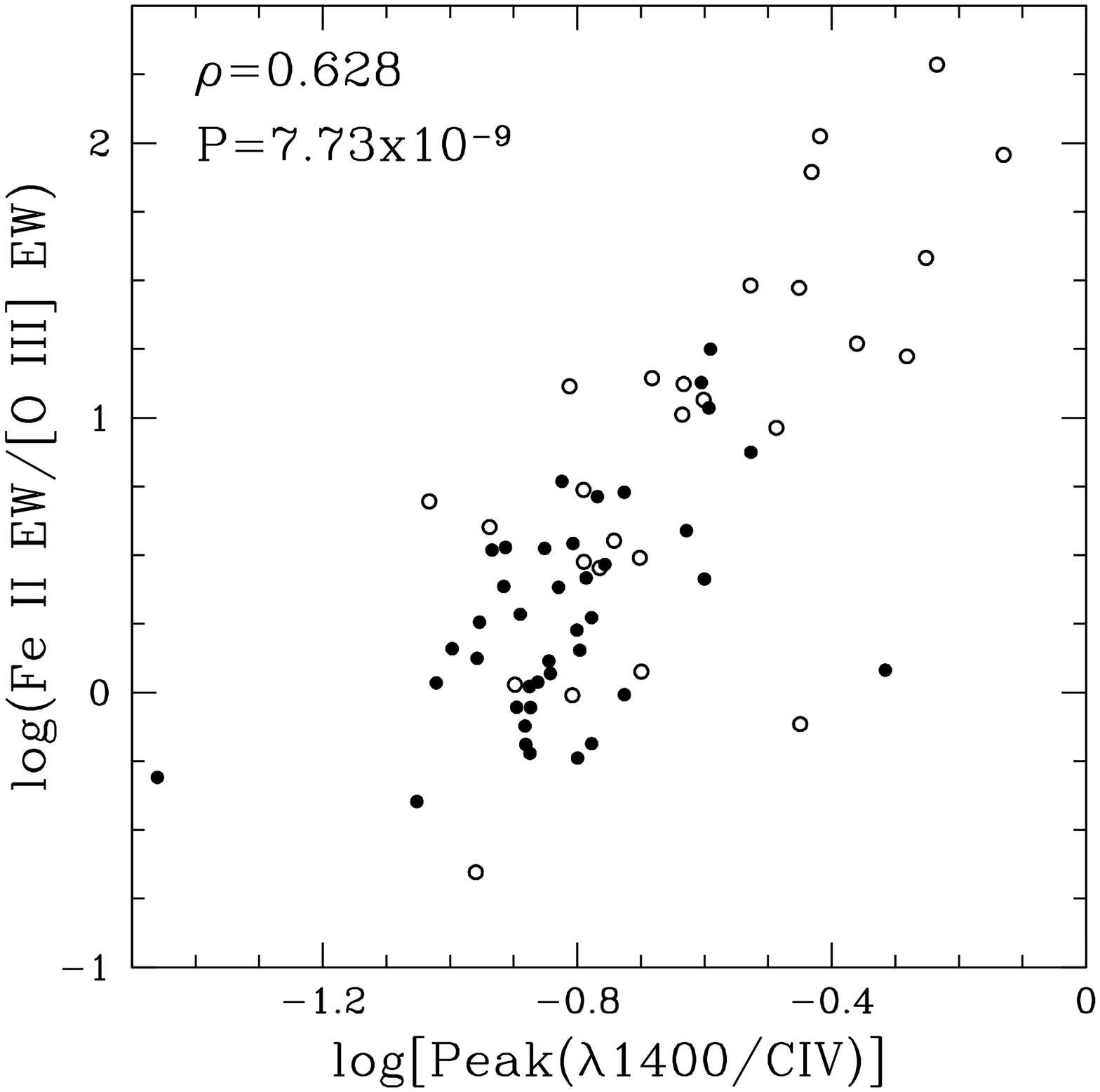}
\end{minipage}
\hspace{0.7cm}
\begin{minipage}[!b]{8cm}
\centering
\includegraphics[width=8cm]{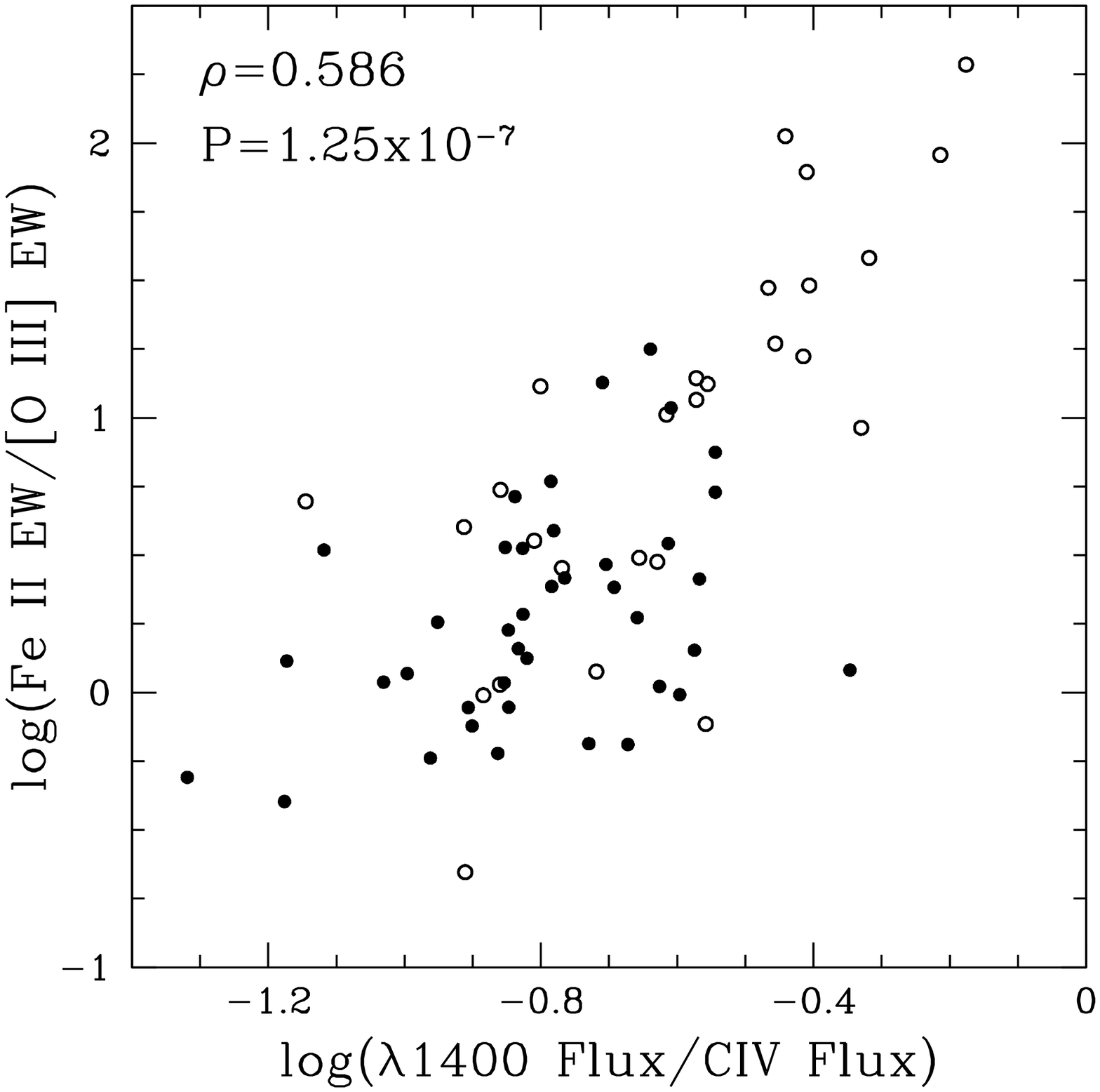}
\end{minipage}
\hspace{0.7cm}
\begin{minipage}[!b]{8cm}
\centering
\includegraphics[width=8cm]{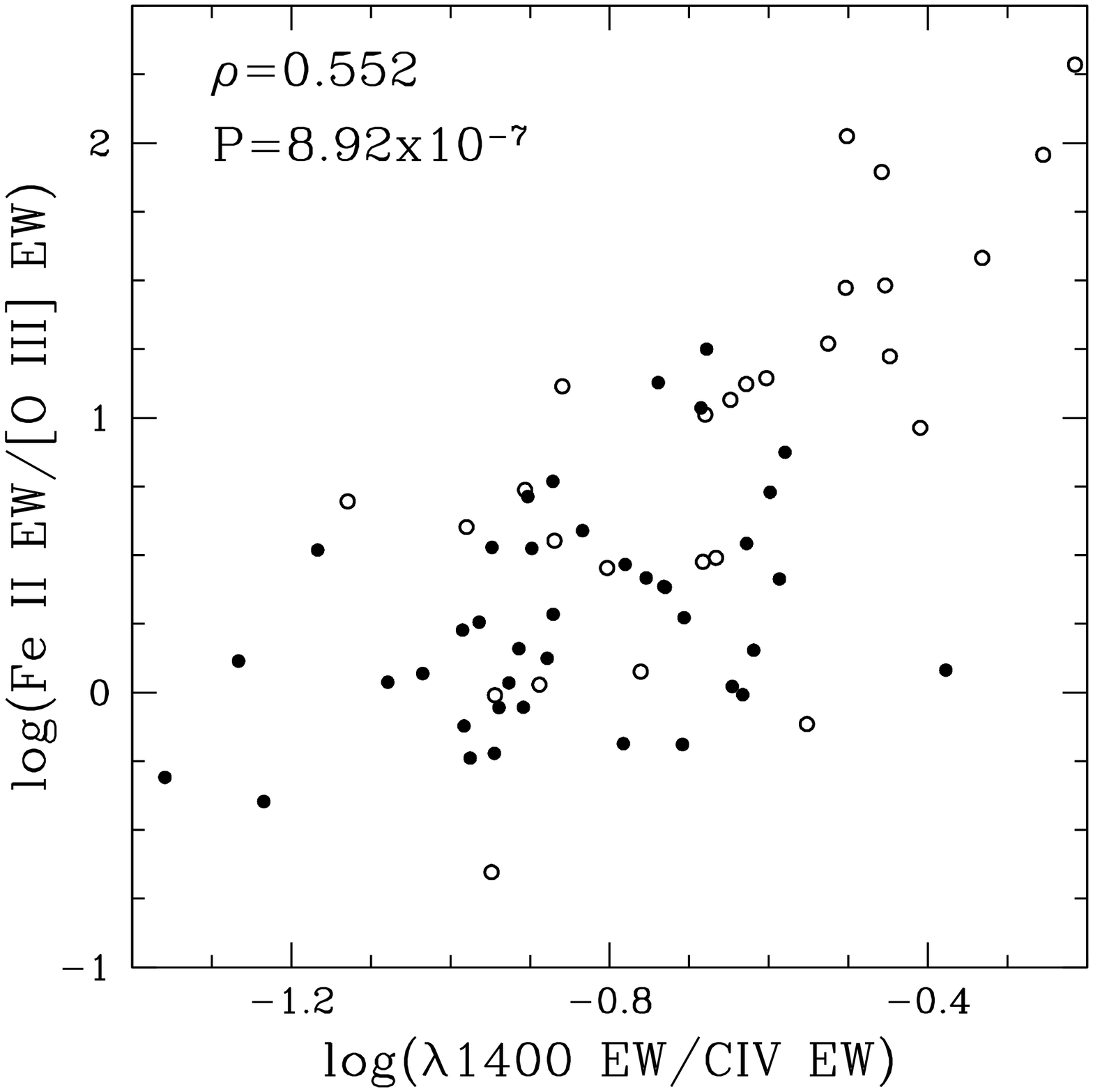}
\end{minipage}     
\hspace{0.7cm}       
\begin{minipage}[!b]{8cm}
\centering
\includegraphics[width=8cm]{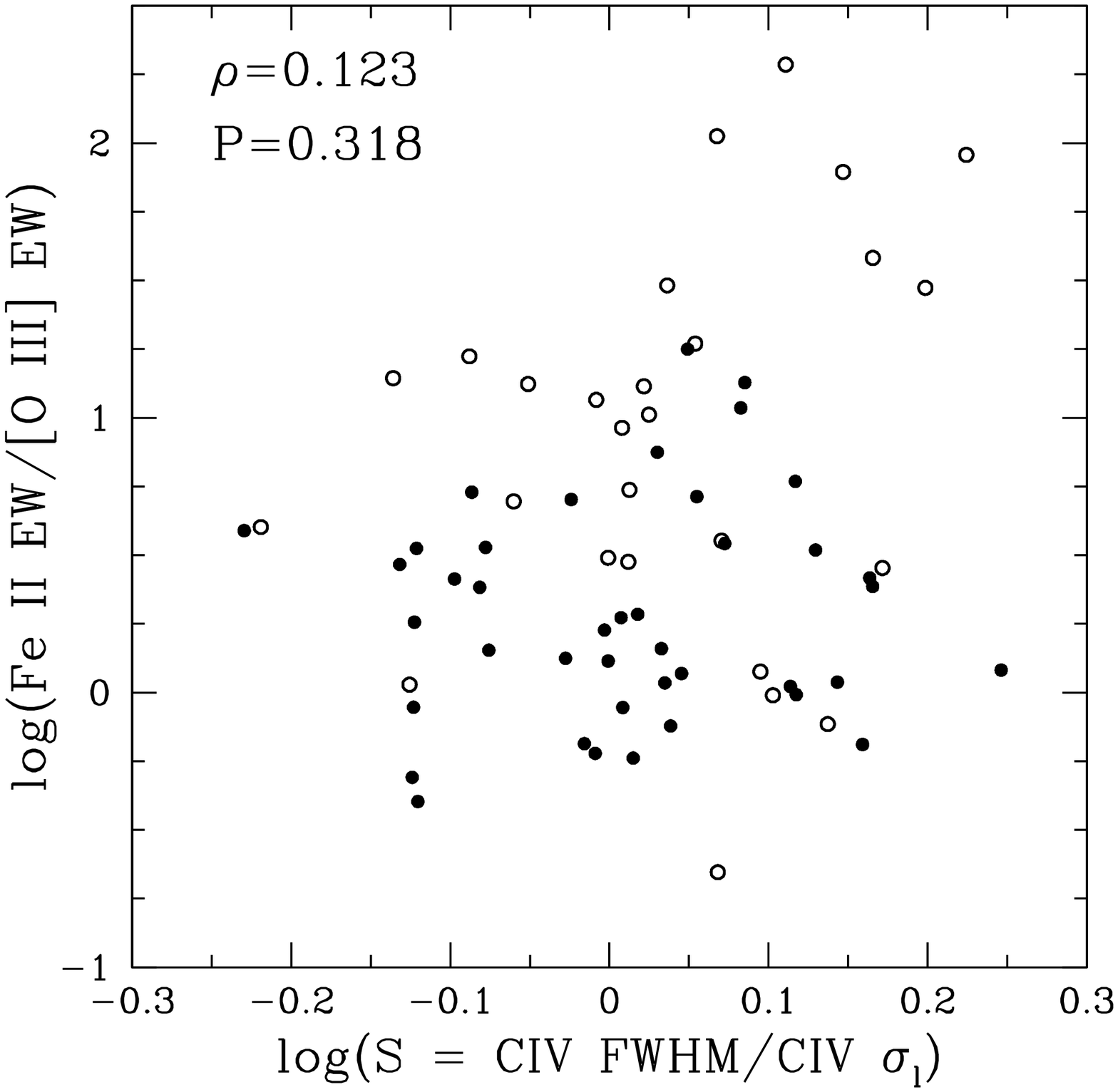}
\end{minipage}                     
\caption{The optical EV1 indicator against four UV EV1 indicators.  The Spearman Rank correlation coefficient ($\rho$) and associated probability ($P$) that the points will be thus distributed by chance is listed in the upper left corner.  The ratio of the $\lambda1400$ to \CIV\ peaks (top left panel) is the best proxy for the optical EV1 indicator.  Radio-loud objects are indicated by solid circles and open circles are radio-quiet.}
\label{fig:EV1}
\end{figure*}

%eigenvector 1 indicators
EV1 is a suite of spectral properties, so in addition to Peak($\lambda$1400/\CIV) and the shape of \CIV\ there are many UV measurements that are strongly correlated with optical EV1 \citep[e.g.,][]{brotherton99a}.  It is not necessary to restrict our analysis to Peak($\lambda$1400/\CIV) if another spectral property yields a better mass correction.  Because of the practical applications of this work, we were particularly interested in EV1 indicators that are simple to measure and can be obtained with limited wavelength coverage.  This rules out UV EV1 indicators like the blueshift of \CIV; without sufficient wavelength coverage obtaining a reliable rest frame for the object and thus establishing the blueshift may be difficult.  We investigated three UV spectral ratios to compare their ability to predict the optical EV1 indicator: the ratios of peak flux, EW, and flux in the line for $\lambda1400$/\CIV.  We found that, of these, Peak($\lambda$1400/\CIV) is most significantly correlated with EW(\FeII/\OIII) (Figure~\ref{fig:EV1}) as well as the FWHM residuals.

%DISCUSSION
%%%%%%%%%%%%%%%%%%%%%%%%%%%%%%%%%%%%%%%%%%%%%%%%%%%%%%%%%%%%%%%%%%%%%%%%%%%%%%%%%
\section{Discussion}
\label{sec:discussion}
\subsection{Data quality and sample selection}
%implications for low SN data
The rehabilitation of black hole mass estimates derived from the FWHM of \CIV\ has significant implications for calculating black hole masses in large numbers of objects.  \citet{denney13} shows that, until now, the most reliable \CIV-based black hole mass estimates required a measurement of line dispersion.  The line dispersion measurement is less sensitive to the presence of a core component in the \CIV\ line, and with a shape correction the contamination can be accounted for to some degree.  However, line dispersion measurements are incredibly sensitive to the information in the wings of the emission line which can be lost in low S/N data.  The low signal to noise of survey-quality data introduces issues for line dispersion measurements, namely that an accurate measurement cannot be obtained.  Using FWHM-based black hole masses and applying the shape correction requires a measure of line dispersion, and thus high-quality data.

The introduction of a peak ratio UV EV1 correction to a \CIV\ FWHM mass provides a robust way of measuring more accurate masses from the UV spectral region.  FWHM measurements do not suffer as much from these S/N constraints, making them ideal for estimating black hole mass in survey-quality data.  A correction to estimate the strength of the core component in the \CIV\ line based on a peak ratio is easy to measure and more robust to low signal to noise, enabling reliable black hole mass estimates for thousands of high-redshift objects in surveys like the Sloan Digital Sky Survey (SDSS).  

\citet{denney09a} performed extensive simulations on the effects of characterizing line widths in data with a variety of S/N ratios.  They found that, in low S/N spectra, the ability to characterize the line width is compromised.  Furthermore, depending on the procedure used to make the measurement, this can result in either over or underestimated line widths, although for FWHM an underestimate is most common.  This discrepancy has the potential to create an offset between \CIV\ and \Hb-based black hole masses that is buried in the black hole mass scaling relations.  If this effect is present, it has been wrapped into the correction for the EV1 bias.  This should be noted, but is not necessarily the most likely outcome or an issue if it has occurred.  According to \citet{denney09a}, the effect of allowing a contaminating narrow, non-virial component in the emission line is larger than S/N effects and \citet{denney12} has verified the presence of a shape bias in black hole estimates based on \CIV.  Finally, for any investigation using spectra similar to ours, the initial offset in the mass residuals from zero exists when applying the \citet{vestergaard06} scaling relations and the improvement still stands whether we attribute it to measurement issues or EV1.  Brotherton et al. (in prep) will perform a more quantitative determination of the amount of improvement that is due exclusively to EV1.

%RM bias
We also note that any C IV-based black hole mass scaling relationship created from a sample that does not have a representative EV1 average will result in biased black hole masses.  This is equivalent to the ``shape bias'' suggested by \citet{denney12}.  Figure~\ref{fig:Mbhcorr} (left) shows that the distribution of points in the plot of \Hb\ versus \CIV\ derived masses using the relationships of \citet{vestergaard06} are not evenly spread around the one-to-one line as expected.  While some of this effect may result from the combination of measurement and S/N bias \citep[e.g.,][]{denney09a}, we also expect a bias since the reverberation sample does not have a representative distribution in EV1, and in particular are deficient in objects at the strong \FeII/weak \OIII\ end.  We will quantitatively investigate this effect in an upcoming paper (Brotherton et al. in prep.).

\begin{figure*}
\begin{minipage}[!b]{8cm}
\centering
\includegraphics[width=8cm]{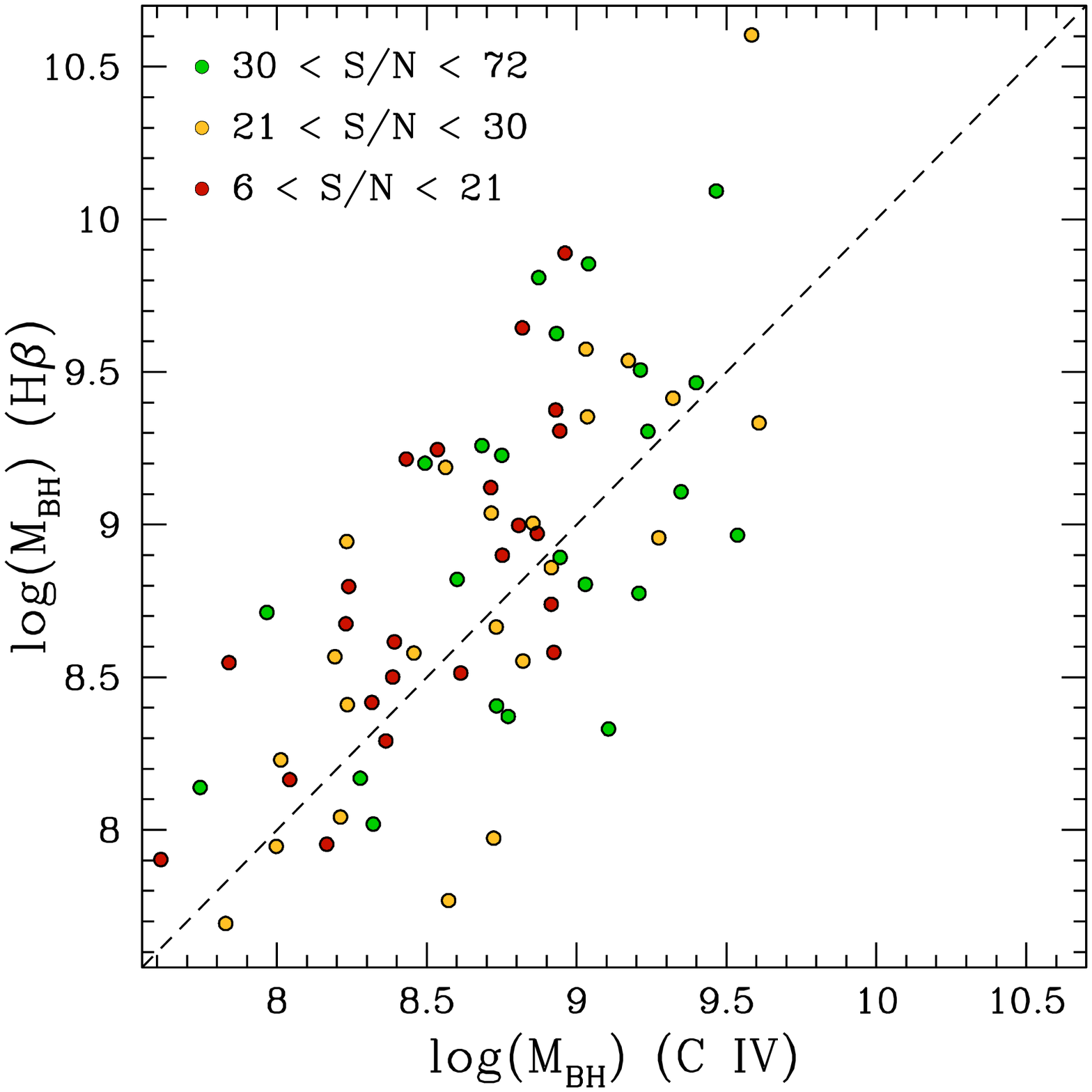}
\end{minipage}\hspace{0.6cm}
\hspace{0.6cm}
\begin{minipage}[!b]{8cm}
\centering
\includegraphics[width=8cm]{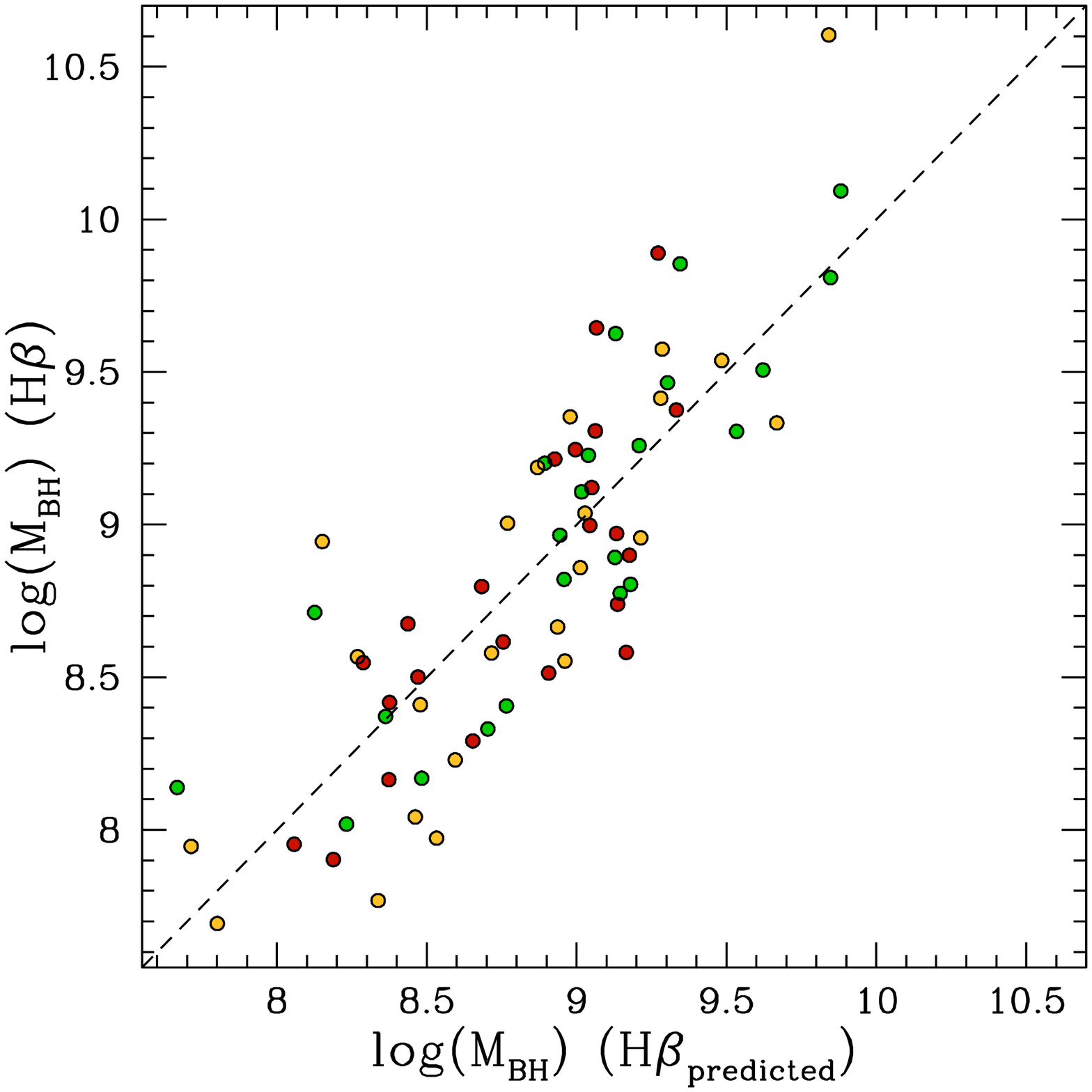}
\end{minipage}
\hspace{0.6cm}                   
\caption{\Hb-based black hole masses versus \CIV-based black hole masses (left panel) and the mass calculated from the predicted \Hb\ based on UV spectral information (right panel) color coded by S/N.  The one-to-one line is shown.  Red points indicate objects with the lowest S/N, yellow points have intermediate S/N, and green points have the highest S/N.}
\label{fig:Mbhcorrsn}
\end{figure*}

In light of S/N concerns for \CIV, we consider whether the success of our black hole correction might depend on the S/N of the spectrum to which it is applied.  Figure~\ref{fig:Mbhcorrsn} is identical to Figure~\ref{fig:Mbhcorr} except that the sources have been color coded into three S/N bins ($6<S/N<21,\,21<S/N<30,\textrm{ and }30<S/N<72$), each with an equal number of objects.  In general, it does not appear that the the sources in any S/N bin are isolated in mass or distance from the one-to-one relationship.  There does seem to be an absence of higher-S/N sources with too-small \CIV\ black hole masses (upper right of the left panel), but this is the opposite of what is expected to result from an S/N bias.  According to \citet{denney09a}, FWHM tends to be underestimated in low-S/N data, thus we might expect an abundance of low-S/N objects in this region of the plot.  There is no trend in the scatter between the \Hb\ masses and the predicted \Hb\ masses in the S/N bins: the final scatter is 0.31, 0.38, and 0.31 dex for the high, medium, and low-S/N bins, respectively.  Given that all of our data are fairly high S/N, above the S/N$\sim$5 cutoff suggested by \citet{denney09a} for FWHM calculations, this result is not surprising.

\subsection{The broad line region}
%orientation
The FWHM of \Hb, which is reliable for probing virialized gas and estimating black hole masses, depends on orientation as indicated by the radio core dominance \citep{wills86}.  The FWHM of \CIV\ shows no such dependence, although \citet{vestergaard02} shows that the full width at one quarter maximum (FWQM) does depend on orientation.  This behavior in the \CIV\ line is likely the result of the two different components of the line having different orientation dependancies.  The virialized component is orientation dependent, as with \Hb, and is targeted by FWQM which probes the base of the line where this component dominates emission.  The non-virial component is independent of orientation and is targeted by FWHM which probes higher in the line where the line width is sensitive to the emission from non-virial gas.  In this scenario, once the FWHM of \CIV\ has been corrected for a non-virial contribution to emission, it should depend on orientation.  An investigation of this prediction is currently underway (Runnoe et al. in prep.)

%ILR
One model that is often invoked to explain the two-component structure of the \CIV\ emission was proposed by \citet{wills93b} and has the BLR is split into two structures, the very broad line region (VBLR) and the intermediate line region (ILR).  \citet{brotherton94} compared VBLR and ILR spectra with the results of photoionization models and found that the VBLR is located nearer to the continuum source and is denser than the ILR.  They predict that \CIV\ should have a strong ILR contributions whereas \Hb\ and the 1400 \AA\ feature will not.  The ILR emission is not consistent with emission from the traditional NLR as the physical parameters required to create the observed emission are different, with the ILR having higher densities \citep{brotherton94} and larger velocity widths \citep{brotherton94,denney12}.  In this model scenario, the non-reverberating, low-velocity core component in the \CIV\ line is attributed to the ILR and creates the peaky \CIV\ line profiles.  \citet{wills93b} posits that this emission may also originate in a biconical outflow, which is not mutually exclusive to a contribution from ILR emission.  This type of multi-component velocity model shows good agreement with observed \CIV\ profiles.  \citet{bachev04} observes a distinct VBLR component in objects with low-redshift, composite spectra with small EW(\FeII/\OIII), although they attribute the low-velocity emission to a traditional NLR region in their decomposition of the line.  The ILR is found to agree with with observed \CIV\ line profiles in the literature \citep[e.g.,][]{wills93b,brotherton94,denney12}. 
%Nevertheless, a two-component structure to the velocity widths does seem to fit well with observed \Hb\ line profiles \citep{hu08,zhu09}.  

%big picture
The goal of this work is to progress in our ability to measure physical properties of quasars with the highest accuracy and precision possible.  Other investigations have already taken steps in this direction, including \citet{runnoe12a} and \citet{assef11} who provide prescriptions for removing an orientation and color bias, respectively, from black hole mass calculations.  These investigations suggest that future progress can be made by determining sources of scatter between black hole mass estimates and including those sources in our measurements of physical parameters.

%CONCLUSION
%%%%%%%%%%%%%%%%%%%%%%%%%%%%%%%%%%%%%%%%%%%%%%%%%%%%%%%%%%%%%%%%%%%%%%%%%%%%%%%%%
\section{Conclusions}
\label{sec:conclusion}
Black hole masses estimated from the \CIV\ emission line have large scatter compared to \Hb-based masses caused by a non-virialized component, consistent with a contribution from the ILR in the \CIV\ line that is weak or absent from \Hb.  The strength of the non-virial component of the \CIV\ line is known to scale with indicators of EV1, allowing it to be separated from the virial \CIV\ emission.  Using the quasi-simultaneous optical and UV spectra from the SED atlas of \citet{shang11}, we investigated methods for using UV spectral information to more reliably predict \Hb-based line widths and black hole masses.  We employed the ratio of the peak flux of the \siivoiv\ line at 1400 \AA\ to the peak flux of \CIV\ as an indicator of the optical EV1, as measured by the equivalent width ratio of \OIII\ to \FeII, and showed that adding this UV EV1 indicator to a measure of \CIV\ line width predicts the \Hb\ line width with increased accuracy.  

We also investigated whether the EV1 dependence persists when line width is calculated using the line dispersion rather than FWHM.  It does, but the effect is much less significant.  Predicting the \Hb-based line width and mass using line dispersion and UV spectral measurements provides only a minimal improvement to using a \CIV-only prescription.  

Including UV EV1 when calculating black hole mass from UV spectra increases agreement between \CIV\ and \Hb-based black hole mass estimates, reducing the scatter between the two from 0.43 dex to 0.33 dex.  Using this prescription, black hole masses can be more reliably calculated from the \CIV\ line using FWHM which is robust to low S/N data.

%ACKNOWLEDGEMENTS
%%%%%%%%%%%%%%%%%%%%%%%%%%%%%%%%%%%%%%%%%%%%%%%%%%%%%%%%%%%%%%%%%%%%%%%%%%%%%%%%%
\section*{Acknowledgments}

J. Runnoe would like to thank Kelly Denney for helpful discussions during the preparation of this work.  Z. Shang acknowledges support by the National Natural Science
Foundation of China through Grant No. 10773006 and Tianjin Distinguished Professor Funds.  We thank the anonymous referee for suggestions that improved this work.

%The authors also appreciate the time of an anonymous referee whose comments improved the paper.

%BIBLIOGRAPHY
%%%%%%%%%%%%%%%%%%%%%%%%%%%%%%%%%%%%%%%%%%%%%%%%%%%%%%%%%%%%%%%%%%%%%%%%%%%%%%%%%
\bibliographystyle{/Users/jrunnoe/Library/texmf/bibtex/bst/mn2e}
\bibliography{./all.051713}
\clearpage

\label{lastpage}
\end{document}